\documentclass[graybox]{svmult}

\usepackage{mathptmx}       % selects Times Roman as basic font
\usepackage{helvet}         % selects Helvetica as sans-serif font
\usepackage{courier}        % selects Courier as typewriter font
\usepackage{type1cm}        % activate if the above 3 fonts are
\usepackage{makeidx}         % allows index generation
\usepackage{graphicx}        % standard LaTeX graphics tool
\usepackage{multicol}        % used for the two-column index
\usepackage[bottom]{footmisc}% places footnotes at page bottom

\begin{document}

\title*{Non-Centrosymmetric Heavy-Fermion Superconductors}
\author{N. Kimura and I. Bonalde}
\institute{Noriaki Kimura \at Center for Low Temperature
Science, Tohoku University, Sendai, Miyagi 980-8578, Japan.
\email{kimura@mail.clts.tohoku.ac.jp} \and Ismardo Bonalde \at
Centro de F\'{\i}sica, Instituto Venezolano de Investigaciones
Cient\'{\i}ficas, Apartado 20632, Caracas 1020-A, Venezuela.
\email{ijbonalde@gmail.com}}
\maketitle

\abstract{In this chapter we discuss the physical properties of
a particular family of non-centrosymmetric superconductors
belonging to the class heavy-fermion compounds. This group
includes the ferromagnet UIr and the antiferromagnets
CeRhSi$_3$, CeIrSi$_3$, CeCoGe$_3$, CeIrGe$_3$ and CePt$_3$Si,
of which all but CePt$_3$Si become superconducting only under
pressure. Each of these superconductors has intriguing and
interesting properties. We first analyze CePt$_3$Si, then
review CeRhSi$_3$, CeIrSi$_3$, CeCoGe$_3$ and CeIrGe$_3$, which
are very similar to each other in their magnetic and electrical
properties, and finally discuss UIr. For each material we
discuss the crystal structure, magnetic order, occurrence of
superconductivity, phase diagram, characteristic parameters,
superconducting properties and pairing states. We present an
overview of the similarities and differences between all these
six compounds at the end.}

\section{CePt$_3$Si}

The enormous interest in superconductors without inversion
symmetry started with the discovery of superconductivity in the
heavy-fermion compound CePt$_3$Si \cite{bauer}, which exhibits
long-range antiferromagnetic (AFM) order below the Neel
temperature $T_{N}=2.2$ K and becomes superconducting at the
critical temperature $T_{c}=0.75$ K. CePt$_3$Si is the only
known heavy-fermion (HF) compound without inversion symmetry
that superconducts at ambient pressure, as opposed to the cases
of CeIrSi$_3$, CeRhSi$_3$, CeCoGe$_3$, CeIrGe$_3$ and UIr. The
heavy-fermion character has been established from the large
Sommerfeld coefficient $\gamma_n \approx 0.39$ J/K$^2$mol
\cite{bauer}. The coexistence of antiferromagnetism and
superconductivity has been proved by zero-field muon-spin
relaxation \cite{amato1} and neutron scattering \cite{metoki}.
Although CePt$_3$Si has been intensively studied both
theoretically and experimentally many questions regarding its
superconducting state remain unresolved, in part because the
observed properties have some sample dependence.  We will focus
here on the superconducting properties of CePt$_3$Si, giving
only a brief review on normal state and magnetic behaviors (a
detailed review of these can be found in
Refs.~\cite{mine8,mine11}).

\subsection{Crystal Structure, Sample Growth and Characteristic Parameters}
\label{subsec:31}

\begin{figure}
% \sidecaption
\centering \scalebox{0.7}{\includegraphics{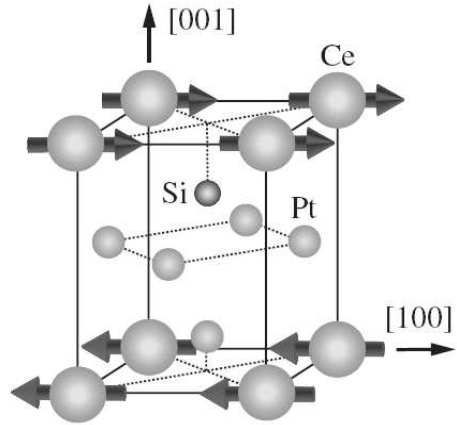}}
 \caption{Crystal and magnetic structures of CePt$_3$Si.}
 \label{f:8}
 \end{figure}

CePt$_3$Si crystallizes in a tetragonal crystal structure with
space group $P4mm$ (No. 99) without inversion symmetry
\cite{bauer}. The lattice parameters are listed in
Table~\ref{tab:2}. The unit cell has one formula unit with one
Ce, one Si and two Pt inequivalent sites. The absence of
inversion symmetry comes from the missing mirror plane
$(0,0,\frac{1}{2})$ (see Fig.~\ref{f:8}). The antiferromagnetic
lattice has an ordered wave vector (0,0,1/2), with a magnetic
moment oriented ferromagnetically along the axis [100] and
antiferromagnetically along the axis [001], as indicated in
Fig.~\ref{f:8}.

%
%--------------- Figure 1 -------------------
%------------- Figure 1 ----------------------
%
%

The melting temperature of CePt$_3$Si is about 1390 $^{\circ}$C
\cite{gribanov}. An isothermal section  of the Ce-Pt-Si phase
diagram at 600 $^{\circ}$C was presented by Gribanov et al.
\cite{gribanov}, who indicated that the interaction of Ce, Pt
and Si leads to the formation of at least nine stable ternary
phases. Seven of these ternary phases have a fixed composition.
Polycrystalline samples of CePt$_3$Si have been grown by argon
arc melting and high-frequency melting and single crystals by
the Bridgman and high-frequency techniques. These samples are
usually  annealed under high vacuum around 900 $^{\circ}$C for
2-3 weeks. Interestingly, the annealing process has been linked
to a Si excess \cite{jskim1}. Growing very high quality single
crystals of CePt$_3$Si has taken a long path that has led to
the resolution of most of the problems in identifying the true
superconducting properties of this compound.

\begin{table}[htb]
\centering \caption{Normal and superconducting parameters of
CePt$_3$Si} 	
\begin{tabular}{p{7cm}p{4cm}}
\hline\noalign{\smallskip}
		Crystal structure                              &   tetragonal  \\
		Space group                                    &   $ P4mm$        \\
		Lattice parameters                              &   $a=4.072$~{\AA} \\
                                    & $c= 5.442$ {\AA} \\
\noalign{\smallskip}\hline\noalign{\smallskip}
		Sommerfeld value of specific heat at $T_c$             &   $\gamma_n = 300-400$~mJ/mol~K$^2$ \\
        Effective electron mass (Fermi sheet $\alpha$) &
        $m^* \sim 11-23 \,m_0$ \\
		Mean free path                                 &   $l = 1200-2700$~{\AA}   \\
		Antiferromagnetic transition temperature       &   $T_{N}  =2.2$~K   \\
		Antiferromagnetic propagation vector           &   $ \vec q = (0,0,1/2)$  \\
        Staggered magnetic moment $\mathbf{m_Q}$ along &   [100]  \\
		Ordered moment per Ce atom                     &   $\mu_s = 0.16~\mu_B$  \\
\noalign{\smallskip}\hline\noalign{\smallskip}
		Superconducting transition temperature         &   $T_{c} = 0.75$~K (or 0.5 K ?)\\
        Specific heat jump at $T_{c}$                 &
        $\Delta C/\gamma_nT_{c} \approx 0.25$ \\
		Upper critical field (small anisotropy)        &   $H_{c2}(0) \sim 3$~T \\
		Thermodynamic critical field                   &   $H_c(0) =  26$~mT \\
		Ginzburg-Landau coherence length               &   $\xi(0) \sim 104$~{\AA}    \\
		Ginzburg-Landau parameter                      &   $\kappa = 82$  \\
		London penetration depth                       &   $\lambda(0) \approx 0.86~\mu$m \\
        Nodal structure                                &   Line
        nodes  \\
\noalign{\smallskip}\hline\noalign{\smallskip}		
\end{tabular} 	
\label{tab:2}
\end{table}

\subsection{Normal State}
\label{subsec:32}

\subsubsection{Phase Diagram and Magnetic Properties}

\begin{figure}
\sidecaption
 \scalebox{0.8}{\includegraphics{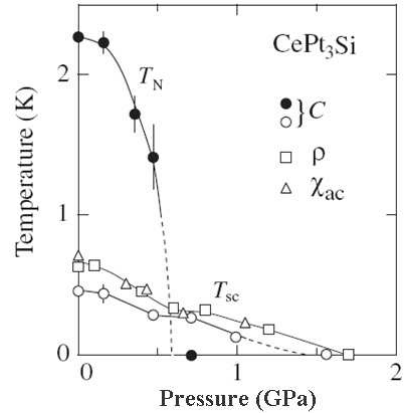}}
 \caption{Temperature-pressure phase diagram of CePt$_3$Si showing the
 coexistence of the
 antiferromagnetic and superconducting phases for pressures below 0.7
 GPa \cite{tateiwa}.}
 \label{f:9}
 \end{figure}

Figure~\ref{f:9} shows the temperature-pressure phase diagram
of CePt$_3$Si determined by specific heat, resistivity and ac
magnetic susceptibility measurements \cite{tateiwa}. The
antiferromagnetic $T_N$ and superconducting $T_{c}$ transition
temperatures decrease with increasing pressure and become zero
around $P_{AF}=0.7 $~GPa and $P_{c}=1.6$~GPa, respectively.

%
%--------------- Figure 2 -------------------
%------------- Figure 2 ----------------------
%
%

\begin{figure}
% \sidecaption
 \scalebox{1.2}{\includegraphics{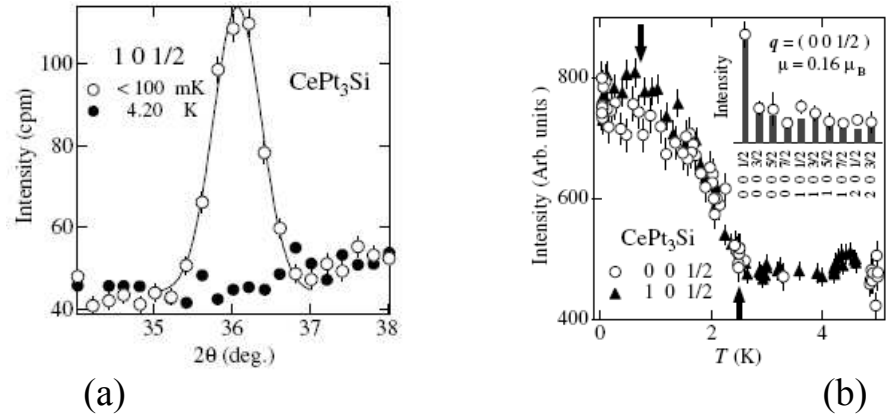}}
 \caption{(a) (101/2) AFM
Bragg reflection observed below 0.1 K (open circles) and the
background measured at 4.2 K (solid circles). (b) The intensity
of (001/2) and (101/2) magnetic reflections as a function of
temperature, shown by open circles and solid triangles,
respectively. Up and down pointing arrows indicate $T_N$ and
$T_c$, respectively \cite{metoki}.}
 \label{f:10}
 \end{figure}

The phase diagram indicates that there are two distinct
superconducting phases: one below $ P_{AF} $ coexisting with
the AFM phase and another above $ P_{AF} $ being presumably the
only ordered phase. The coexistence of the superconducting and
AFM phases was confirmed by neutron-scattering measurements
that clearly show two superlattice peaks below and above the
superconducting critical temperature (see Fig.~\ref{f:10})
\cite{metoki}. The observed peaks (0\,0\,1/2) and (1\,0\,1/2)
correspond to an AFM vector $Q_0=(0\,0\,1/2)$. The magnetic
structure consists of ferromagnetic sheets of rather small Ce
moments of $0.16~\mu_B$/Ce stacked antiferromagnetically along
the $c$ axis (see Fig.\ref{f:8}). Here, $\mu_B$ is the Bohr
magneton. The small value of the moment relative to
2.54~$\mu_B$/Ce of  Ce$^{3+}$ may partially be explained
through the itinerant character of Ce 4$f$-electrons involved
in the formation of the heavy quasiparticles. In parts the
reduction of the moment is also due to the Kondo screening
effect viewing these electrons as almost localized moments
\cite{bauer}. In general, the magnetic response of CePt$_3$Si
in the normal state involves also the interplay of crystal
electric field splitting of the 4$f$-orbitals and Kondo
interaction (see, for example, Ref.~\cite{mine11}).

%
%--------------- Figure 3 -------------------
%------------- Figure 3 ----------------------
%
%

\subsection{Superconducting State}

At ambient pressure superconductivity in CePt$_3$Si appears
within a Fermi-liquid state, as evidenced by quantum
oscillations \cite{hashimoto1} and resistivity measurements
\cite{bauer,yasuda}. However, CePt$_3$Si becomes a
non-Fermi-liquid under pressure, as indicated by the linear
temperature behavior of the resistivity above 0.4 GPa
\cite{yasuda}.

Most of the unusual superconducting properties found initially
in CePt$_3$Si have been clarified by now, but others have
appeared. First results, like second anomalies or a small peak
just below the superconducting transition in the NMR $1/T_1T$,
are not observed in the latest measurements carried out on new
single crystals. These early results were associated with
sample dependence: sample preparation, off-stoichiometry,
impurity phases and/or annealing conditions. The new puzzling
feature is a transition temperature that falls sometimes below
0.5 K. \cite{motoyama1,mukuda2,mine13,mine14}.

Several theoretical approaches have been developed to try to
understand this superconductor
\cite{samokhin,sergienko,frigeri,fujimoto3,yanase1}. The models
take into account the splitting of the spin-degenerate bands
caused by the absence of inversion symmetry. Antiferromagnetic
order effects in the heavy-fermion superconductors are also
considered \cite{fujimoto3,yanase1}. Even though much
experimental and theoretical efforts have been dedicated to
clarify its physics, CePt$_3$Si continues to be the most
interesting and challenging of all superconductors without
spatial inversion symmetry.

\subsubsection{Probing the Pairing Symmetry}
\label{pairingsys}

Several techniques have been employed to test the Cooper
pairing state in non-centrosymmetric CePt$_3$Si. We will review
what has been done thus far.

\paragraph{\bf Spin State}

\begin{figure}
% \sidecaption
 \scalebox{0.9}{\includegraphics{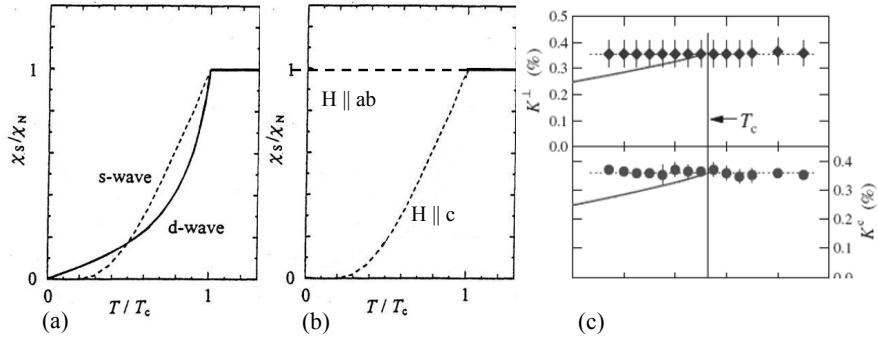}}
 \caption{Electronic spin susceptibility expected in (a) spin-singlet states
 $s$-wave and $d$-wave and (b) spin-triplet states. (c)
 Experimental electronic spin susceptibility of CePt$_3$Si showing no change across the
 superconducting transition for all orientations of the applied magnetic field (upper panel: $ \vec{H} \perp \hat{z} $ and lower panel $ \vec{H} \parallel \hat{z} $)  \cite{yogi2}.}
 \label{f:15}
 \end{figure}

One of the most direct probes of the spin state of the pairing
is the ratio of the superconducting to the normal electron-spin
paramagnetic susceptibility, $\chi_s$ and $\chi_n$,
respectively, which for a spin-singlet pairing state may be
written as
%
%--------------- Eq 1 -------------------
\begin{equation}
\label{suscept} \frac{\chi_s}{\chi_n}=-2 \int_0^\infty d\xi
\left\langle \frac{\partial f(E)}{\partial E} \right\rangle_{\hat{k}} \, .
\end{equation}
%------------- Eq 1 ----------------------
%
The integration is over $\xi$, the energy of the free electrons
relative to the Fermi level, $f$ is the Fermi distribution
function, and $E=\sqrt{\xi^2 + \Delta(\hat{k})^2}$ is the
energy of the quasiparticles. The bracket $ \langle \cdots
\rangle_{\hat{k}} $ denotes the angular average. Here, $\Delta
(\hat{k}) $ is the energy gap that in general depends on the
momentum direction $\hat{k}$ and temperature. For $T << T_c$
the ratio $\chi_s/\chi_n$ goes as
$(1/\sqrt{T})\exp(-\Delta_0/k_BT)$ for an $s$-wave pairing
state with an isotropic gap and as $T^2$ for a $d$-wave pairing
state where the gap has line nodes (Fig.~\ref{f:15}(a)).
$\Delta_0$ is the zero-temperature value of $\Delta$ and $k_B$
is the Boltzmann constant.

For spin-triplet pairing the ratio $ \chi_s / \chi_n $ is more
complicated and depends on the field orientation. In the most
simple cases, the susceptibility does not change across the
transition if the field is perpendicular to the $\textbf{d}$
vector denoting the spin-triplet gap function, but decreases
continuously down to zero at $ T =0 $ if the applied field is
parallel to the $\textbf{d}$ vector (Fig.~\ref{f:15}(b)).

%
%--------------- Figure 2 -------------------
%------------- Figure 2 ----------------------
%
%
In CePt$_3$Si the experimental result of $\chi_s/\chi_n$ is
puzzling: $ \chi_s/\chi_n $ does not change at all in the
superconducting phase for any orientation of the field as shown
in Fig.~\ref{f:15}(c) \cite{yogi2}. Model calculations
including a sizable ASOC characteristic for the
non-centrosymmetric CePt$_3$Si predict that for both
spin-singlet and spin-triplet states $\chi_s(0)/\chi_n
\rightarrow 1$ for fields parallel to the $z$ axis and
$\chi_s(0)/\chi_n \rightarrow 1/2$ for fields perpendicular to
$z$ \cite{frigeri3,samokhin3}. On the other hand, if electron
correlations are included, $\chi_s/\chi_n$ can be calculated to
be constant across the transition independently of the field
orientation \cite{fujimoto3,yanase1}. Thus, in CePt$_3$Si
spin-susceptibility measurements are not quite useful to
distinguish between spin-singlet and spin-triplet pairings.

The upper critical field $H_{c2}$ can also be used to get
information about the spin configuration of a pairing state. A
magnetic field induces pair breaking via paramagnetic and
orbital mechanisms. The Pauli paramagnetic limiting field $H_P$
can be estimated by comparing the (zero-field) superconducting
condensation energy with the Zeeman energy
\begin{equation}
\frac{1}{2}(\chi_n-\chi_s)H_P^2=\frac{1}{2}N_0\Delta_0^2 \,,
\end{equation}
\noindent where $N_0$ is the density of states at the Fermi
energy. The Pauli susceptibility $\chi_n$ is given by $\chi_n=(g\mu_{B})^2N_0/2$,
where $g$ is the gyromagnetic ratio. $H_P$ is then derived as
\begin{equation}
H_P= \frac{\sqrt{2}\Delta_0}{
g\mu_{B}\sqrt{1-\chi_s/\chi_n}} \,. \label{eq:pauli}
\end{equation}
\noindent As discussed above, for spin-singlet superconductors
$\chi_s$ goes to zero as $T\to 0$. Then, using the BCS value
$\Delta_0=1.76k_{\rm B}T_c$ for a weak-coupling superconductor
and $g=2$ for free electrons, we obtain the well-known estimate
\begin{equation}
H_P^{BCS}=H_P(0)=1.86 T_c \, {\rm [T/K]}  \,. \label{eq:hpbcs}
\end{equation}
This expression is also valid for spin-triplet superconductors
if the field is applied parallel to the $\textbf{d}$ vector. On
the other hand, if the field is applied perpendicular to the
$\textbf{d}$ vector $\chi_s=\chi_n$ as $T\rightarrow 0$ and
Eq.~(\ref{eq:pauli}) yields $H_P\to \infty$. This would imply
the absence of the Pauli paramagnetic limiting effect for
fields along the $ab$ plane.

\begin{figure}
\sidecaption
 \scalebox{0.51}{\includegraphics{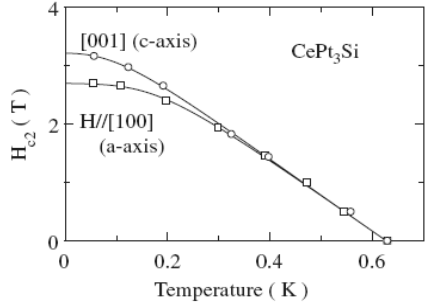}}
 \caption{Temperature dependence of the upper critical field in CePt$_3$Si
 showing a weakly anisotropic behavior \cite{yasuda}.}
 \label{f:16}
 \end{figure}

Measurements on high-quality single crystals of CePt$_3$Si show
a weak anisotropy for the upper critical field $H_{c2}(0)$ with
a value around 3 T (Fig.~\ref{f:16}) \cite{yasuda} that exceeds
the standard BCS weak-coupling paramagnetic limit
$H_P(0)\approx 1$ T.

\vspace{8pt}
%
%--------------- Figure 2 -------------------
%------------- Figure 2 ----------------------
%

From Eq.~(\ref{eq:pauli}) and the predictions for
$\chi_s/\chi_n$ in the non-centrosymmetric superconductors, no
limiting behavior is expected for fields parallel to the $c$
axis, whereas $H_P(0)=\Delta_0/\mu_B \approx 1.4$ T for fields
in the $ab$ plane. As mentioned above, modifications of $
\chi_s $ due to correlation effects as well as magnetic
ordering could eliminate the paramagnetic limiting for all
field directions \cite{fujimoto3,yanase1}. It was also
suggested that the realization of a so-called helical phase for
fields perpendicular to the $c$ axis would strongly reduce
paramagnetic limiting effects \cite{kaur,samokhin4}.

The orbital limiting field $H_{orb}$ is expressed by
\begin{equation}
H_{orb}(T)=\frac{\Phi_0}{2\pi\xi^2(T)} \,, \label{eq:orb}
\end{equation}
where $\Phi_0$ is the flux quantum. $H_{orb}(T=0)$ can be in
principle obtained by using the BCS expression
$\xi(0)=0.18\hbar v_{\rm F}/(k_{\rm B}T_c)$, where $v_{\rm F}$
is the Fermi velocity. However, $H_{orb}(T=0)$ is usually
estimated from the formula \cite{helfand}
\begin{equation}
H_{orb}(T)= h(T) H_{c2}^{\prime}T_c \,.\label{eq:initial}
\end{equation}
\noindent Here, $H_{c2}^{\prime}\equiv -dH_{c2}/dT|_{T=T_c}$.
$h(0)=0.727$ for weak-coupling BCS superconductors in the clean
limit. Using the data of Fig.~\ref{f:16} we estimate
$H_{c2}^{\prime} \approx -6.8$ T/K and from
Eq.~(\ref{eq:initial}) $H_{orb}^{BCS}\approx 3.7$ T. This is
about the value of $H_{c2}(0)$ for both field orientations,
which suggests that CePt$_3$Si may be restricted by the orbital
depairing limit. In such a case the spin-triplet state should
be favorable. Both paramagnetic and orbital mechanisms will be
discussed in more detail in Sect. 2.

\paragraph{\bf Nodal Structure}

The structure of the energy gap is directly related to the
symmetry of the Cooper pairing. The energy gap is isotropic for
$s$-wave spin-singlet superconductors and usually has zeroes
(nodes) for other symmetries. Thus, the determination of the
presence of nodes in the energy gap is crucial to establish
pairing with symmetries lower than the $s$-wave. The existence
of nodes in the energy gap leads to low-temperature power laws
($~T^n$) in several superconducting properties, instead of the
BCS exponential temperature response observed for an
isotropically gapped excitation spectrum. In CePt$_3$Si
magnetic penetration-depth, thermal-conductivity and
specific-heat measurements show power-law behaviors indicative
of line nodes in the energy gap.

\begin{figure}
% \sidecaption
 \scalebox{0.85}{\includegraphics{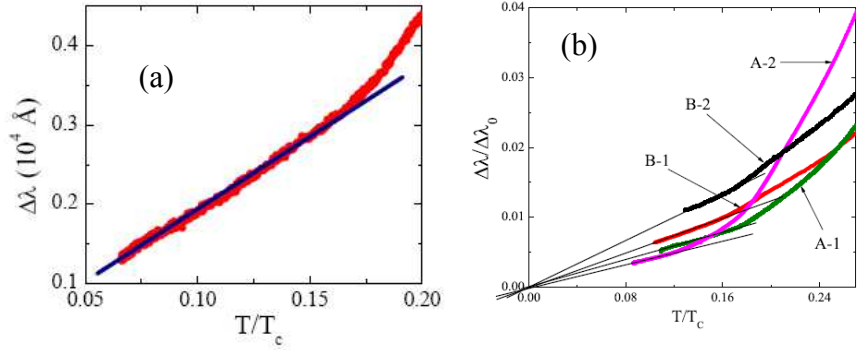}}
 \caption{Penetration depth in (a) a polycrystalline sample and (b) single crystals of CePt$_3$Si.
The linear temperature behavior indicates line nodes in the
energy gap. The single crystals used in the penetration-depth
measurements shown in (b) have different defect concentrations,
which suggests that the linear behavior is unaffected by
disorder \cite{mine13}.}
 \label{f:17}
 \end{figure}

A linear temperature dependence of the magnetic penetration
depth $\lambda(T)$ below 0.16 $T_c$ was first found in a
polycrystalline sample of CePt$_3$Si (Fig.~\ref{f:17}(a))
\cite{mine7}, and later also in single crystals \cite{mine13}.
In the local limit of the electrodynamics of superconductors,
the magnetic penetration depth is given by
%
%--------------- Eq 1 -------------------
\begin{equation}
\label{supden}
\left[\frac{\lambda^2(0)}{\lambda^2(T)}\right]_{ij}=\frac{n^s_{ij}(T)}{n}=3\left\langle
\hat{k}_i\hat{k}_j\left[ 1-\int d\xi\left( \frac{-\partial
f}{\partial E_{\textbf{k}}}\right) \right] \right\rangle_{\hat{k}} \, .
\end{equation}
%------------- Eq 1 ----------------------
%
\noindent In superconductors with inversion symmetry $ \Delta
\lambda(T) \propto T$ is expected in the low-temperature limit
for an energy gap with line nodes \cite{mineev}. Thus, the
linear response in CePt$_3$Si was taken as evidence for line
nodes. Surprisingly, the temperature response in the
penetration depth is not affected by the sample quality, as
opposed to what occurs in unconventional superconductors with
line nodes in the gap \cite{mine13}. Figure~\ref{f:17}(b)
displays the low-temperature region of the penetration depth of
several single crystals of different quality. A clear linear
behavior is observed in all of them. We note that the
superfluid density $\rho_s(T)$ shows a small anisotropy
\cite{mine9}, in agreement with the upper critical field
result.

%
%--------------- Figure 2 -------------------
%------------- Figure 2 ----------------------
%
%

\begin{figure}[t]
% \sidecaption
 \scalebox{0.7}{\includegraphics{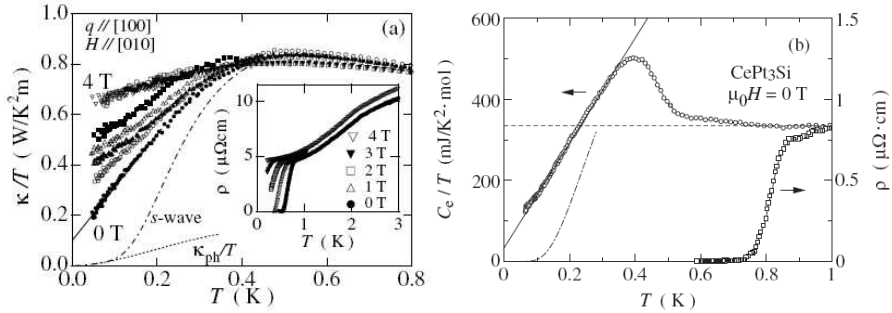}}
 \caption{Temperature dependence of (a) thermal conductivity \cite{izawa} and
 (b) specific heat \cite{takeuchi2} of CePt$_3$Si. The linear temperature
 response suggests line nodes in the energy gap.}
 \label{f:18}
 \end{figure}

Thermal transport measurements also suggest the presence of
line nodes by showing a residual linear term in $\kappa(T)/T$
as $T\rightarrow0$ (Fig.~\ref{f:18}(a)). In superconductors
with inversion symmetry such a linear term is expected when the
energy gap has nodes, and is due to impurity scattering. The
quasiparticle thermal conductivity has universal components in
the low-temperature limit ($k_BT<\gamma$)
%
%--------------- Eq 1 -------------------
\begin{equation}
\label{thercond} \kappa_{ii}=\frac{\pi^2}{3}N_0\nu_F^2 T
\left\langle \hat{k}_i \hat{k}_i
\frac{\gamma^2}{(\gamma^2 + \Delta_{\textbf{k}}^2)^{3/2}} \right\rangle_{\hat{k}} \,
\end{equation}
%------------- Eq 1 ----------------------
%
that are linear functions of temperature at low $T$ and whose
proportionality constants depend on the specific form of the
order parameter \cite{mineev}. $\gamma$ is the quasiparticle
decay rate. In CePt$_3$Si the experimental $\kappa(T)/T= 0.1$
W/(K$^2 \cdot $m) is in good agreement with the calculated
universal conductivity limit 0.09 W/(K$^2 \cdot$m)
\cite{izawa}. Moreover, the $H$ dependence of $\kappa$ follows
the prediction by a theory of Doppler-shifted quasiparticles in
a superconductor with line nodes \cite{izawa,kubert}.

%
%--------------- Figure 2 -------------------
%------------- Figure 2 ----------------------
%
%

In the low-temperature limit the electronic specific heat of
CePt$_3$Si has been found to follow the expression $C_{el}/T =
A + BT$, with $A= 34.1$ mJ/(K$^2 \cdot$ mol) and $B=1290$
mJ/(K$^3 \cdot$ mol) (Fig.~\ref{f:18}(b)) \cite{takeuchi2}. In
general, the electronic specific heat is given by \cite{mineev}
%
%--------------- Eq 1 -------------------
\begin{equation}
\label{thercond} C_{el}=2N_0 \int_{-\infty}^{\infty} d\xi
\left\langle  E_{\textbf{k}} \frac{\partial f}{\partial
E_{\textbf{k}}} \right\rangle_{\hat{k}}  \, .
\end{equation}
%------------- Eq 1 ----------------------
%
For superconductors with inversion symmetry, in the
low-temperature limit $C_{el} \propto T^2$ for a gap with line
nodes.  The residual linear term in $C_{el}/T$ of CePt$_3$Si is
considered to be caused by impurities or by electrons on part
of the Fermi surface that do not participate in
superconducting. Thus, the observed behavior of the electronic
specific heat is taken as evidence for line nodes in the energy
gap \cite{takeuchi2}.

For a theoretical discussion on line nodes in
non-centrosymmetric superconductors, see other chapters in this
book.

\section{Ce$TX_3$ Compounds}

\subsection{Crystal Structure and Related Compounds}
\begin{figure}[tb]
%\sidecaption
	\includegraphics[width=0.9\textwidth]{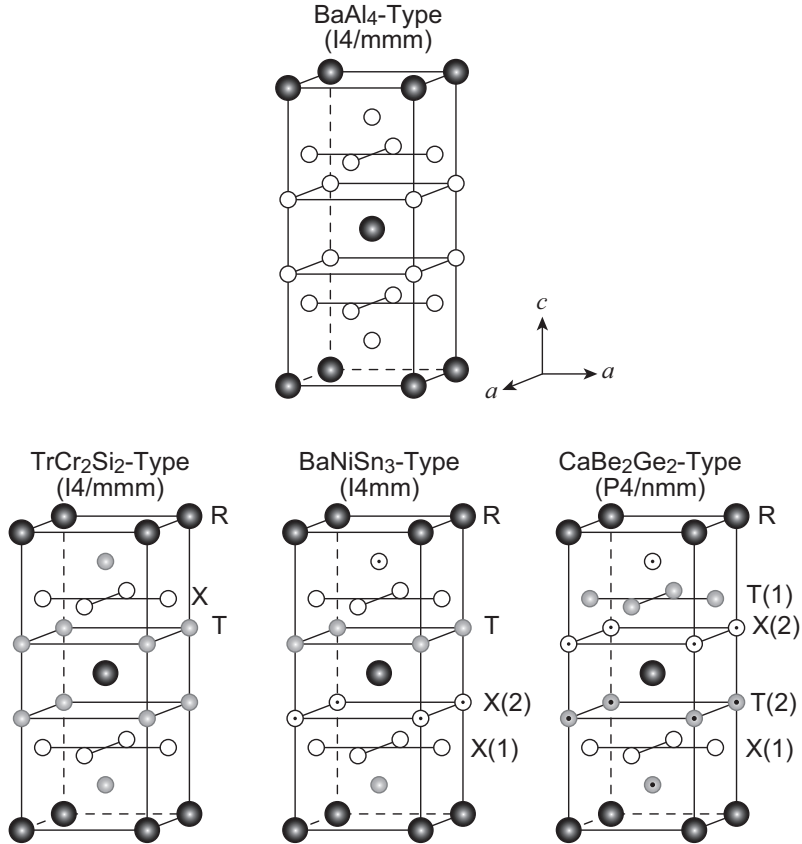}
\caption{BaAl$_4$-type crystal structure and its three
derivatives TrCr$_2$Si$_2$, BaNiSn$_3$ and CaBe$_2$Ge$_2$. Only
the BaNiSn$_3$-type structure does not have an inversion
center. } \label{fig:crystal}
\end{figure}

Most Ce$TX_3$ compounds crystallize in the BaNiSn$_3$-type
tetragonal structure with space group $I4mm$ (No.107)
\cite{lejay}. The BaNiSn$_3$-type structure derives from the
BaAl$_4$-type structure whose basic frame is the body-centered
tetragonal lattice shown at the top in Fig.~\ref{fig:crystal}.
There are two other derivatives of the BaAl$_4$-type structure,
the ThCr$_2$Si$_2$ and CaBe$_2$Ge$_2$ types. Some heavy-fermion
superconductors crystallize into the former structure: e.g.
CeCu$_2$Si$_2$ \cite{steglich}, CeCu$_2$Ge$_2$ \cite{jaccard},
CePd$_2$Si$_2$ \cite{mathur}, CeRh$_2$Si$_2$ \cite{movshovich}
and URu$_2$Si$_2$ \cite{palstra}. The latter structure is often
found in $R$Pt$_2X_2$ compounds \cite{szytula}, where $R$
denotes a rare-earth element. The ThCr$_2$Si$_2$- and
CaBe$_2$Ge$_2$-type structures have an inversion center, while
the BaNiSn$_3$-type structure does not. Fig.~\ref{fig:crystal}
displays the BaAl$_4$-type crystal lattice and its three
derivatives.

%%%%%%%%%%%%%%%%% FIGURE 1 %%%%%%%%%%%%%%%%%%%
%%%%%%%%%%%%%%%%% FIGURE 1 %%%%%%%%%%%%%%%%%%%

The atomic framework of the BaNiSn$_3$-type structure can be
alternatively displayed as a sequence of planes of the same
atoms $R-T-X(1)-X(2)-R-T-X(1)-X(2)-R$ along the $c$ axis of the
tetragonal structure, where $T$ and $X$ denote a transition
metal and Si/Ge, respectively. The point group of the
BaNiSn$_3$-type structure is C$_{4v}$, which lacks the mirror
plane and a two-fold axis normal to the $c$ axis ($z$ axis).
Therefore, a Rashba-like spin-orbit coupling exists in this
system, as it does in CePt$_3$Si that also belongs to C$_{4v}$
\cite{bauer}.

Of the Ce$TX_3$ compounds, the series Ce$T$Si$_3$ ($T$=Co, Ru,
Rh, Pd, Os, Ir and Pt) and Ce$T$Ge$_3$ ($T$=Fe, Co, Rh and Ir)
are known to crystallize in the BaNiSn$_3$-type structure.
Among these, CeRhSi$_3$ \cite{kimura1}, CeIrSi$_3$
\cite{sugitani}, CeCoGe$_3$ \cite{settai3} and CeIrGe$_3$
\cite{honda} have been found to be pressure-induced
superconductors. Single crystals of CeRhSi$_3$ and CeIrSi$_3$,
as well as of CeRuSi$_3$ \cite{kawai2}, can be obtained by the
Czochralski pulling method in a tetra-arc furnace using a
pulling speed of 10 mm/h or 15 mm/h. In these compounds
annealing treatments at 900 $^{\circ}$C, in vacuum, for a week
are usually very effective. Notably, for CeRhSi$_3$ the use of
stoichiometric amounts of the components sometimes yields a
different crystal; e.g., CeRhSi$_2$. To obtain a crystal of
CeRhSi$_3$ an off-stoichiometric composition, typically
Ce:Rh:Si=1:1:3.3, works better. Single crystals of CeCoGe$_3$
are obtained by the Bi-flux method (the Czochralski method is
unsuitable) \cite{thamizhavel}. In this procedure,
arc-melt-prepared ingots of CeCoGe$_3$ and Bi are placed in an
alumina crucible and heated up to 1050 $^{\circ}$C in an argon
atmosphere. After keeping this temperature for a day, the
crucible is cooled down to 650 $^{\circ}$C over a period of two
weeks and then down to room temperature rapidly. Single
crystals of CeRhGe$_3$ can also be obtained by the Bi-flux
technique \cite{kawai2}. Single crystals of CeIrGe$_3$ are
grown by the Bi- and Sn-flux methods \cite{honda,kawai2}.

Other Ce$TX_3$ compounds with BaNiSn$_3$-type structure are
Ce$T$Al$_3$ ($T$=Cu, Au) and CeAuGa$_3$. They order
magnetically at low temperatures \cite{mock, paschen, mentink,
kontani, sugawara}. In particular, CeCuAl$_3$ and CeAuAl$_3$
are suggested to be HF compounds with AFM ground states
\cite{paschen} and have the potential to become
non-centrosymmetric HF superconductors. There are at least two
Ce$TX_3$ compounds that do not have the BaNiSn$_3$-type crystal
lattice: CeNiGe$_3$, which has an orthorhombic structure and
exhibits superconductivity under pressure \cite{nakashima}, and
CeRuGe$_3$, which is reported to have either orthorhombic
\cite{morozkin} or cubic structure \cite{ghosh}.

On the other hand, no uranium-based compound (U$TX_3$) has been
fully confirmed to have the BaNiSn$_3$-type structure. Only
UNiGa$_3$, that exhibits AFM order at 39 K, seems to have this
structure \cite{takabatake}.

Last, we briefly comment on non-heavy-fermion La$TX_3$
compounds. Some of them, like LaRhSi$_3$, LaIrSi$_3$ and
LaPdSi$_3$, also have BaNiSn$_3$-type structures and show
superconductivity at the critical temperature 1.9, 0.9 and 2.6
K, respectively \cite{lejay,settai1,muroD}. In these materials,
however, no evidence for unconventional behavior has been
observed \cite{settai1}.

%%%%%%%%%%%%%%%%%%%%%%%%%%%%%%%%%%%%%%%%%%%%%%%%%%%%%%%%%%%%%
\subsection{Normal State}
%%%%%%%%%%%%%%%%%%%%%%%%%%%%%%%%%%%%%%%%%%%%%%%%%%%%%%%%%%%%%
\subsubsection{Magnetic Properties}
The Ce$TX_3$ compounds exhibit various magnetic ground states
as summarized in Table~\ref{t1}. The magnetic ground states
vary from AFM to intermediate valence (IV) through HF states
with decreasing unit-cell volume $V$. For example, CeCoGe$_3$
($V=183.3$ \AA $^3$) displays an AFM ground state
\cite{thamizhavel, pecharsky, das}, while CeCoSi$_3$ ($V=163.6$
\AA $^3$) is thought to be an IV compound \cite{rupp}.

\begin{figure}[tb]
%\sidecaption
	\includegraphics[width=0.6\textwidth]{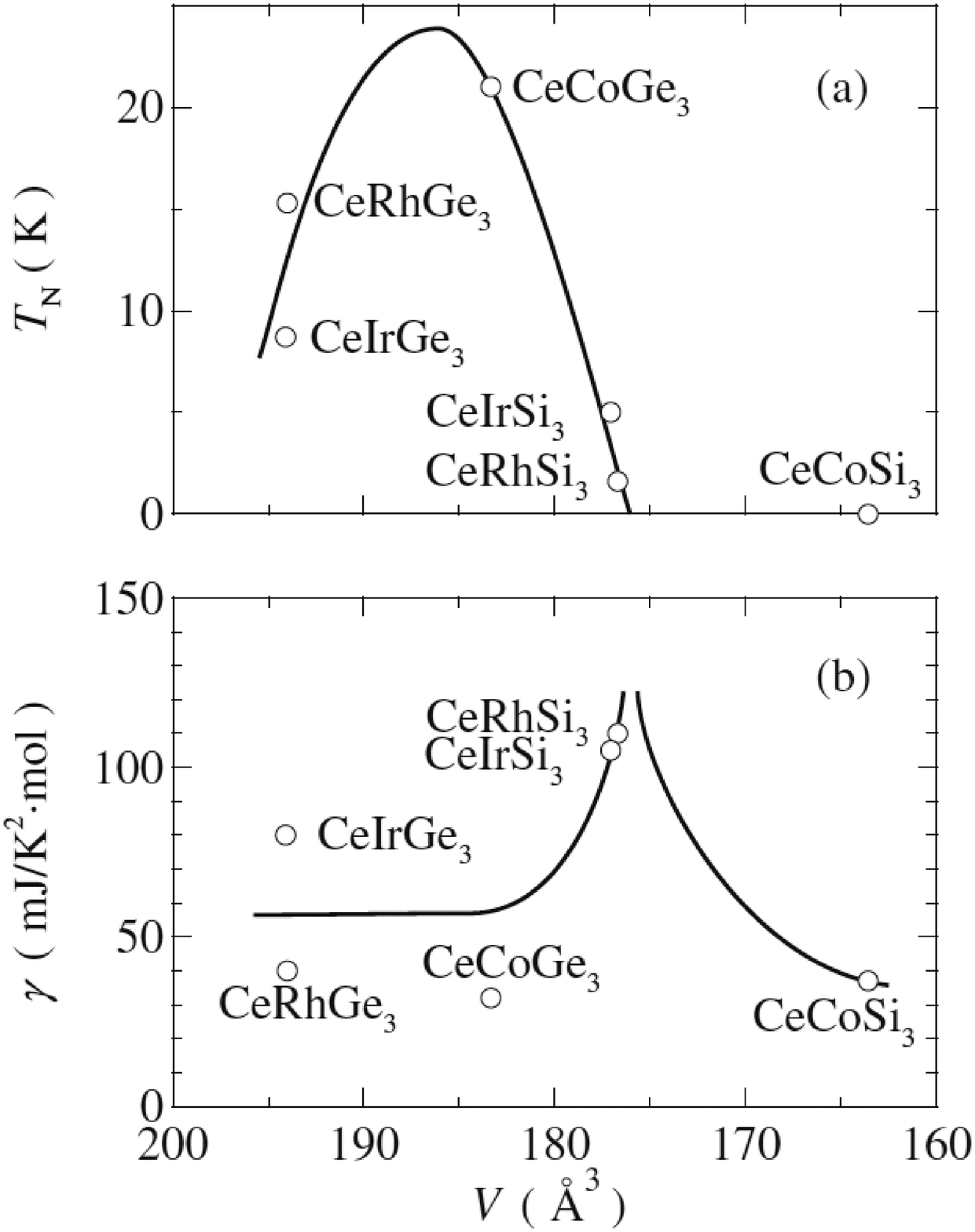}
\caption{Unit-cell volume dependence of (a) the N\'eel
temperature and (b) the $\gamma_n$ value in Ce$T$Si$_3$ and
Ce$T$Ge$_3$ ($T$: Co, Rh, Ir) \cite{kawai2}.}
\label{fig:doniach}
\end{figure}

Figure \ref{fig:doniach} shows the N\'eel temperature $T_{\rm
N}$ and the electronic specific-heat coefficient $\gamma_n$ as
a function of unit-cell volume for Ce$TX_3$ in which $T$
belongs to the Group 9 (Co, Rh and Ir) in the Periodic Table
\cite{kawai2}. $T_N$ approximately follows a simple curve which
peaks at 186 \AA$^3$. This behavior supports the Doniach model
in which the on-site Kondo effect dominates over the inter-site
RKKY interaction with the coupling constant $J$ being
effectively enhanced relative to the kinetic energy with
decreasing unit-cell volume \cite{kawai2}. $\gamma_n$ is also
described by a simple curve which peaks at the unit-cell volume
176 \AA$^3$ at which $T_N$ goes to zero, suggesting that the
$\gamma_n$ value is enhanced by the magnetic fluctuation
arising at the corresponding volume.

%%%%%%%%%%%%%%%%% TABLE I %%%%%%%%%%%%%%%%%%%%
\begin{table}
\caption{Unit-cell volume ($V$), magnetic ground state,
electronic specific-heat coefficient $\gamma_n$, ordering
temperature ($T_N$), Weiss temperature ($\Theta_p$) and
effective moment ($\mu_{eff}$) for the BaNiSn$_3$-type Ce$TX_3$
compounds. The abbreviations PM and AFM denote paramagnetic and
antiferromagnetic ground states, respectively. The
abbreviations IV and HF denote intermediate-valence and
heavy-fermion states, respectively. We consider compounds with
$\gamma_n >100$ mJ/mol$\cdot$K$^2$ to be heavy-fermion
systems.} \label{t1}
\begin{tabular}{lccccccccl}
\hline\noalign{\smallskip}
Compound & $a$ & $c$ & $V$        & Magnetism & $\gamma_n$      & $T_N$ & $\Theta_p$ & $\mu_{eff}$     & Ref.\\
  & [\AA] & [\AA] & [\AA $^3$] &  & [mJ/mol$\cdot$K$^2$] & [K]        & [K]        & [$\mu_B$] &     \\
\noalign{\smallskip}\svhline\noalign{\smallskip} *Ce$T$Si$_3$
& &        &       &         &      &      &      &      &
\\ \noalign{\smallskip}
CeCoSi$_3$ & 4.135   & 9.567  & 163.6  & PM(IV)  & 37   & -    & -840 & 2.80 & \cite{eom} \\
CeRuSi$_3$ & 4.21577 & 9.9271 & 176.43 & PM(IV)  &      & -    &      &      & \cite{kawai2} \\
CeRhSi$_3$ & 4.269   & 9.738  & 177.5  & AFM(HF) & 110  & 1.6  & -128 & 2.65 & \cite{muro,muroD} \\
           & 4.237   & 9.785  & 175.7  &         &      &      &      &      & \cite{kawai3} \\
CePdSi$_3$ & 4.33    & 9.631  & 180.6  & AFM     & 57   & 5.2/3  & -26  & 2.56 & \cite{kitagawa,muroD} \\
CeOsSi$_3$ &         &        &        & PM(IV)  &      &      &      &      & \cite{haen} \\
CeIrSi$_3$ & 4.252   & 9.715  & 175.6  & AFM(HF) & 120  & 5.0  & -142 & 2.48 & \cite{muro,muroD} \\
CePtSi$_3$ & 4.3215  & 9.6075 & 179.42 & AFM     &  29  &
4.8/2.4 &   &      & \cite{kawai1} \\ \noalign{\smallskip}
\noalign{\smallskip}\hline\noalign{\smallskip} *Ce$T$Ge$_3$ & &
&        &         &      &      &      &      &
\\ \noalign{\smallskip}
CeFeGe$_3$ & 4.332   & 9.955  & 186.8  & PM(HF)  & 150  & -    & -90  & 2.6  & \cite{yamamoto} \\
           & 4.3371  & 9.9542 & 187.24 &         &      &      &      &      & \cite{kawaiD}\\
CeCoGe$_3$ & 4.320   & 9.835  & 183.5  & AFM     & 32   & 21/19 & -51 & 2.54 & \cite{eom} \\
           & 4.319   & 9.829  & 183.3  &         &      & 21/12/8 & &      & \cite{thamizhavel} \\
CeRhGe$_3$ & 4.402   & 9.993  & 193.6  & AFM     & 40   & 14.6/10/0.55 & -28 & 2.53 & \cite{muro} \\
           & 4.3976  & 10.0322& 194.01 &         &      & 14.9/8.2 &  &      & \cite{kawai2} \\
CeIrGe$_3$ & 4.409   & 10.032 & 195.0  & AFM     & 80   & 8.7/4.7/0.7 & -21 & 2.39 & \cite{muro} \\
           & 4.401   & 10.024 & 194.2  &         &      & 8.7/4.8  &  &      & \cite{kawai2}\\
\noalign{\smallskip}\hline\noalign{\smallskip}
\end{tabular}
\end{table}
%%%%%%%%%%%%%%%%% TABLE I %%%%%%%%%%%%%%%%%%%%

%%%%%%%%%%%%%%%%% FIGURE 2 %%%%%%%%%%%%%%%%%%%
%%%%%%%%%%%%%%%%% FIGURE 2 %%%%%%%%%%%%%%%%%%%
%

Besides the Group 9 (Co, Rh, Ir) compounds, one can consider
CeFeGe$_3$ to be a potential superconductor because its
$\gamma_n$ is comparable to those of CeRhSi$_3$ and CeIrSi$_3$.
However, in CeFeGe$_3$ superconductivity has not been observed
down to 0.05 K \cite{yamamoto}.

\begin{figure}[tb]
%\sidecaption
	\includegraphics[width=0.92\textwidth]{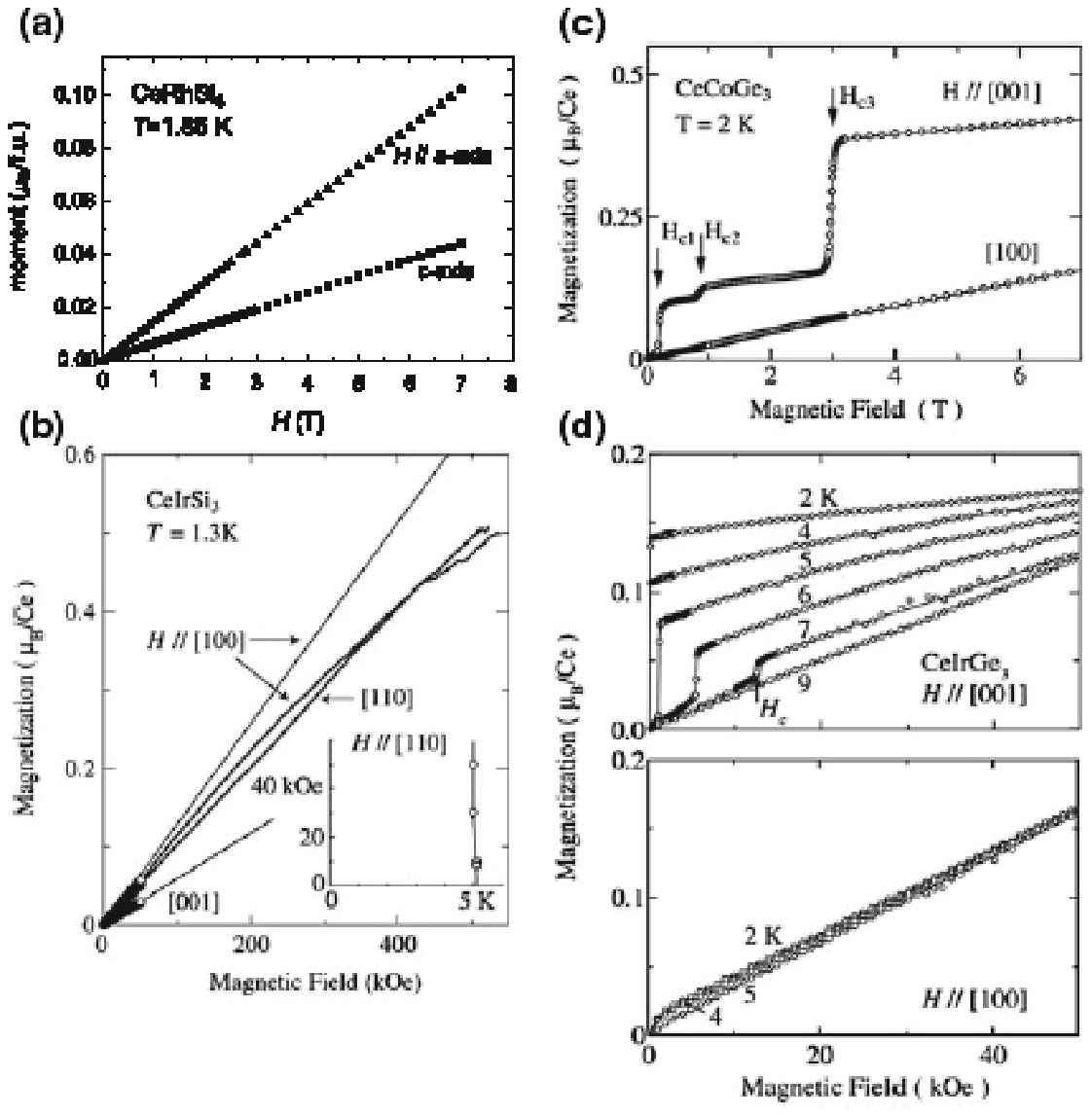}
\caption{Magnetization curves of CeRhSi$_3$, CeIrSi$_3$,
CeCoGe$_3$, and CeIrGe$_3$ \cite{thamizhavel,muroD,okuda}. }
\label{fig:mag_all}
\end{figure}

\begin{figure}[tb]
%\sidecaption
	\includegraphics[width=0.92\textwidth]{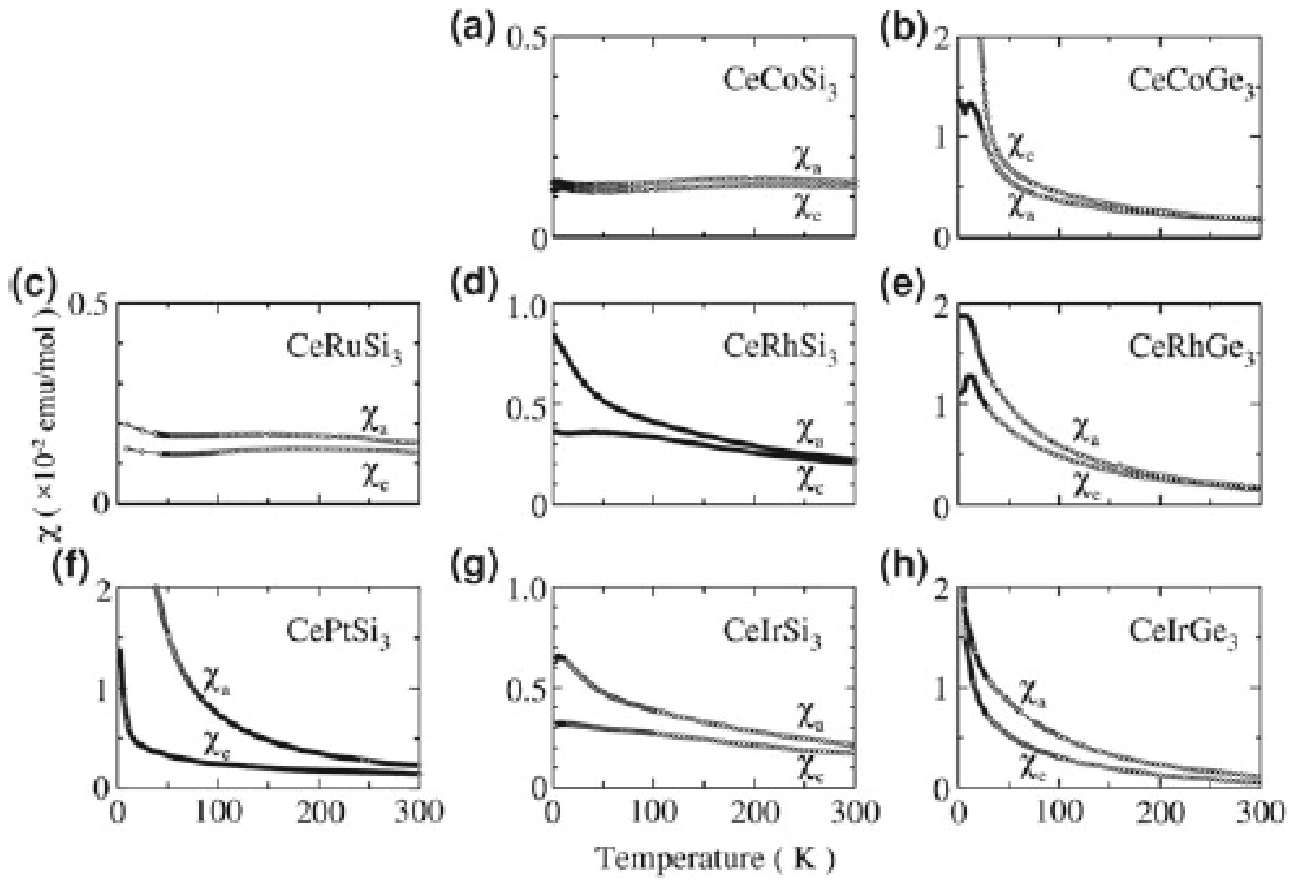}
\caption{Magnetic susceptibility of Ce$TX_3$ compounds as a
function of temperature \cite{kawai2}. } \label{fig:sus_all}
\end{figure}

%-----------------------------------------------------------
\vspace{\baselineskip}
\noindent\textbf{CeRhSi$_3$}\\
As shown in Figs.~\ref{fig:mag_all}(a) and
\ref{fig:sus_all}(d), the magnetic properties of CeRhSi$_3$ are
anisotropic especially at low temperatures \cite{muroD,muro3}.
The induced magnetization along the easy axis [100] has a quite
small value of 0.1$ \mu_B$ at 7 T. The magnetic susceptibility
curves for $H$ parallel to the $a$ and $c$ axes show a strong
anisotropy at low temperatures, while they obey the Curie-Weiss
law above about 150 K. The effective moments $\mu_{eff}$ for
both field directions are $2.65 \mu_B$, which is close to the
expected value for the Ce$^{3+}$ ion. The Weiss temperatures
are negative and very large ($-112$ and $-160$ K for $H$
parallel to the $a$ and $c$ axes, respectively), as often found
for IV compounds. The susceptibility for $H\parallel c$ has a
broad peak around 50 K that is characteristic of HF compounds.
%%%%%%%%%%%%%%%%% FIGURE 3 %%%%%%%%%%%%%%%%%%%
%%%%%%%%%%%%%%%%% FIGURE 3 %%%%%%%%%%%%%%%%%%%
%%%%%%%%%%%%%%%%% FIGURE 4 %%%%%%%%%%%%%%%%%%%
%%%%%%%%%%%%%%%%% FIGURE 4 %%%%%%%%%%%%%%%%%%%

\begin{figure}[tb]
%\sidecaption
	\includegraphics[width=0.6\textwidth]{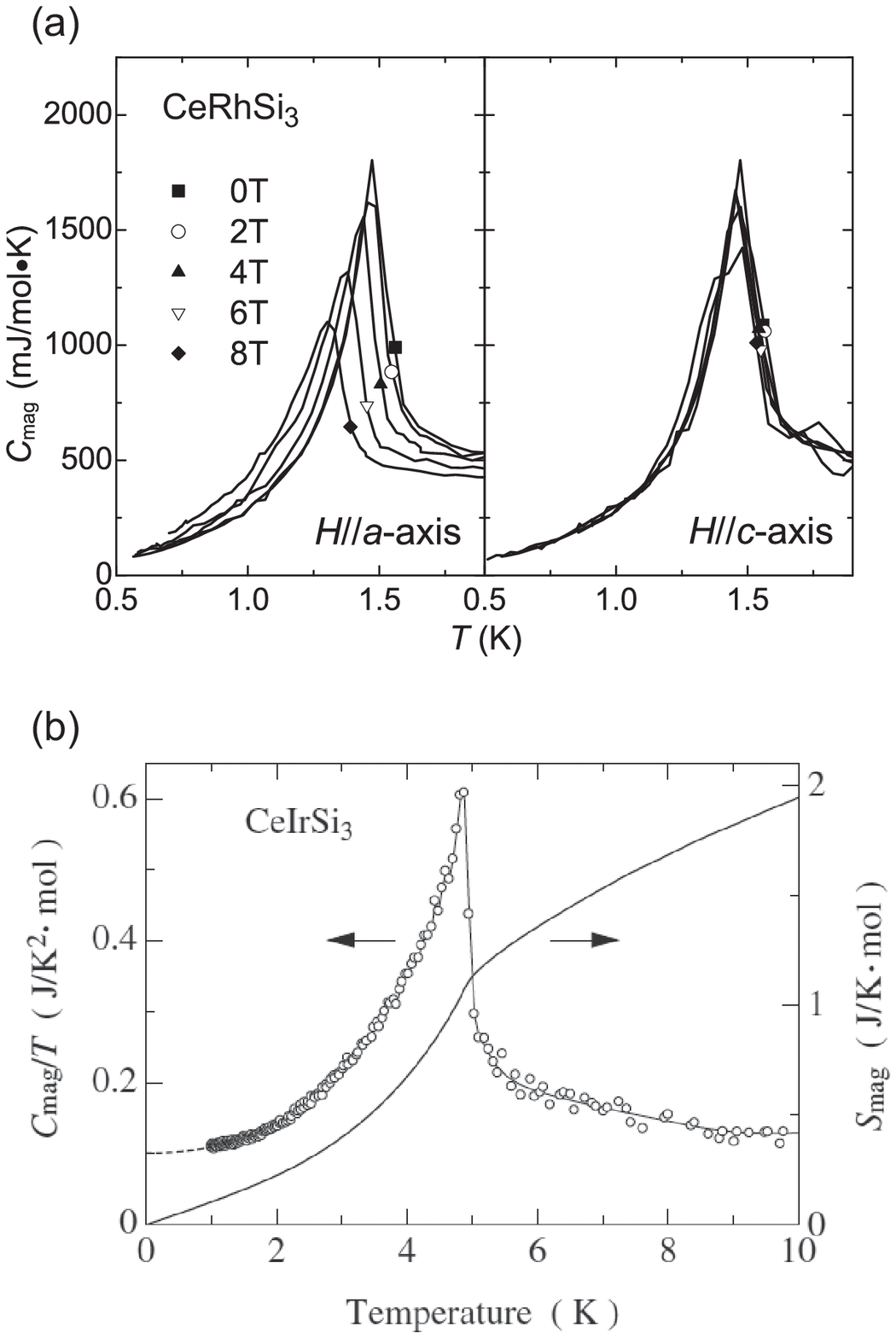}
\caption{(a) Magnetic contribution to the specific heat
$C_{mag}$ of CeRhSi$_3$ as a function of temperature for fields
along the $a$ and $c$ axes \cite{muro3}. (b) $C_{mag}/T$ and
the entropy $S$ of CeIrSi$_3$ as a function of temperature
\cite{okuda}. } \label{fig:C_norm}
\end{figure}

From the jump of the specific heat at $T_N$, the magnetic
entropy gain is estimated to be only 12\% of $R\ln2$
\cite{muro}. The AFM state is robust against a magnetic field
(Fig.~\ref{fig:C_norm}(a)) and survives up to 8 T, although
such a strong field should be sufficient to suppress an AFM
state with a low $T_N$ of the order of 1 K. The temperature and
height of the specific-heat peak decrease with increasing field
along the easy axis ($H\parallel a$). The magnetic contribution
to the specific heat when the field is aligned with the hard
axis ($H\parallel c$) does not change even at 8 T \cite{muro3}.
The tiny entropy gain and the insensitivity to a magnetic field
are attributed to the strong Kondo screening of the $4f$
electron. The Kondo temperature is estimated to be about 50 K
\cite{muro}.
%%%%%%%%%%%%%%%%% FIGURE 5 %%%%%%%%%%%%%%%%%%%
%%%%%%%%%%%%%%%%% FIGURE 5 %%%%%%%%%%%%%%%%%%%

\begin{figure}[tb]
\sidecaption 	
\includegraphics[width=0.5\textwidth]{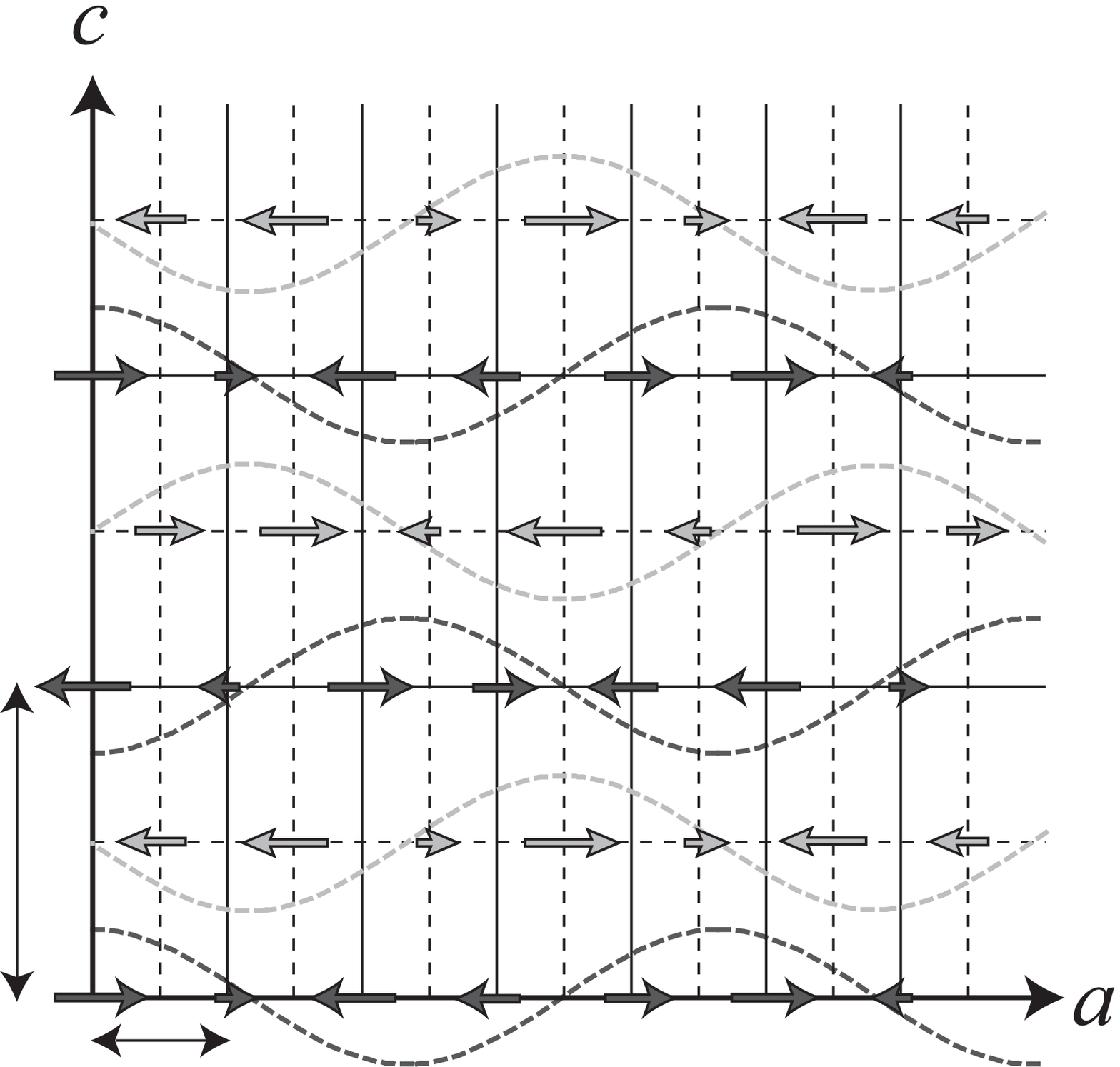}
\caption{Magnetic structure of CeRhSi$_3$. Only the Ce sites at
(0, 0, 0) and (0.5, 0.5, 0.5) positions are projected on the
$ac$ plane. The arrows are depicted to show the size and the
direction of the magnetic moment at the Ce sites \cite{aso2}. }
\label{fig:lsdw}
\end{figure}

The magnetic structure at ambient pressure is revealed by
neutron experiments to be a longitudinal spin-density-wave
(LSDW) type characterized by the propagation vectors $Q=(\pm
0.215,0,0.5)$ with polarization along the $a^*$ axis
\cite{aso}. The magnetic structure is shown in
Fig.~\ref{fig:lsdw}. The staggered moment 0.13 $\mu_B$ is quite
small \cite{aso2}. The incommensurate LSDW structure with such a
strongly suppressed moment suggests that itinerant-electron
magnetism is realized in CeRhSi$_3$. The 4{\em f} electrons are
expected to be strongly hybridized with the conduction
electrons through the Kondo effect, leading to the formation of
the SDW state caused by nesting of the Fermi surface.

It is not obvious whether this magnetic structure persists
under pressure. In the pressure-dependent specific-heat curve,
a shoulder-like transition is seen below $T_N$ at a pressure of
0.55 GPa \cite{tomioka}. The origin of the transition is
unclear at present. Considering that multiple magnetic
transitions are observed in other magnetic Ce$TX_3$ compounds,
the magnetic structure realized in CeRhSi$_3$ at ambient
pressure may change under pressure where superconductivity
appears.
%%%%%%%%%%%%%%%%% FIGURE 6 %%%%%%%%%%%%%%%%%%%
%%%%%%%%%%%%%%%%% FIGURE 6 %%%%%%%%%%%%%%%%%%%

The electrical resistivity below $T_N$ can be fitted by an
antiferro-magnon model \cite{kimura4}. At sufficiently low
temperatures ($T<0.6$ K), the resistivity follows the
Fermi-liquid description of $\rho(T)=\rho_0+AT^2$, where
$\rho_0$ is the residual resistivity. $A=0.19$
$\mu\Omega\cdot$cm/K$^2$ for the current $J$ along the $a$ axis
and $A=0.24$ $\mu\Omega\cdot$cm/K$^2$ for $J\parallel c$. The
ratios $A/\gamma_n^2=1.6\times 10^{-5}$ ($J\parallel a$),
$2.0\times 10^{-5}$ ($J\parallel c$)
$\mu\Omega\cdot$cm$\cdot$K$^2\cdot$mol$\cdot$mJ$^{-2}$ are
close to $1\times 10^{-5}$
$\mu\Omega\cdot$cm$\cdot$K$^2\cdot$mol$\cdot$mJ$^{-2}$ as given
by the Kadowaki-Woods relation \cite{kawo}.

%-----------------------------------------------------------
\vspace{\baselineskip}
\noindent\textbf{CeIrSi$_3$}\\
Although the magnetic structure of CeIrSi$_3$ is unknown at
present, the specific heat, magnetic susceptibility and
electrical resistivity are similar to those of CeRhSi$_3$. The
magnetization curve is anisotropic and $a$ is the easy axis
(Fig.~\ref{fig:mag_all}(b)) \cite{okuda}. The magnetization
increases almost linearly with magnetic field. Neither a
saturation behavior nor metamagnetic transitions are observed
for fields in the basal plane up to 50 T. The induced
magnetization is very small and comparable to that of
CeRhSi$_3$; 0.1$\mu_B$ at 10 T for the easy axis. The
temperature dependence of the magnetic susceptibility
 is anisotropic at low temperatures
(Fig.~\ref{fig:sus_all}(g)) \cite{sugitani,okuda}.

The entropy gain associated with the AFM transition is small,
0.2$R\ln 2$, as in CeRhSi$_3$ (Fig.~\ref{fig:C_norm}(b))
\cite{okuda}. The AFM state is robust against a magnetic field,
as shown in the inset of Fig.~\ref{fig:mag_all}(b). The
electrical resistivity below $T_N$ can be fitted by an
antiferro-magnon model as done in CeRhSi$_3$. The coefficient
of the $T^2$ term $A=0.04$ $\mu\Omega\cdot$cm/K$^2$ for
$J\parallel c$ and $J\perp c$ \cite{okuda} is about one fifth
of the coefficients of CeRhSi$_3$. The ratio
$A/\gamma_n^2=3.6\times 10^{-6}$
$\mu\Omega\cdot$cm$\cdot$K$^2\cdot$mol$\cdot$mJ$^{-2}$ is much
smaller than the ratios in CeRhSi$_3$ but still in the range of
the Kadowaki-Woods relation.

%-----------------------------------------------------------
\vspace{\baselineskip}
\noindent\textbf{CeCoGe$_3$}\\
Unlike CeRhSi$_3$ and CeIrSi$_3$, CeCoGe$_3$ exhibits three
successive AFM transitions \cite{thamizhavel}. Correspondingly,
the magnetization for $H\parallel c$ shows three-step
metamagnetic transitions. It reaches $M_s/4$, $M_s/3$ and
$M_s$, where $M_s=$0.43$\mu_B$/Ce, at each transition. The
anisotropic magnetization curve indicates an Ising-like
magnetism with the easy axis along the $c$ axis, which is
different from what is found in CeRhSi$_3$ and CeIrSi$_3$ as
seen in Fig.~\ref{fig:sus_all}.

\begin{figure}[b]
\sidecaption 	
\includegraphics[width=0.23\textwidth]{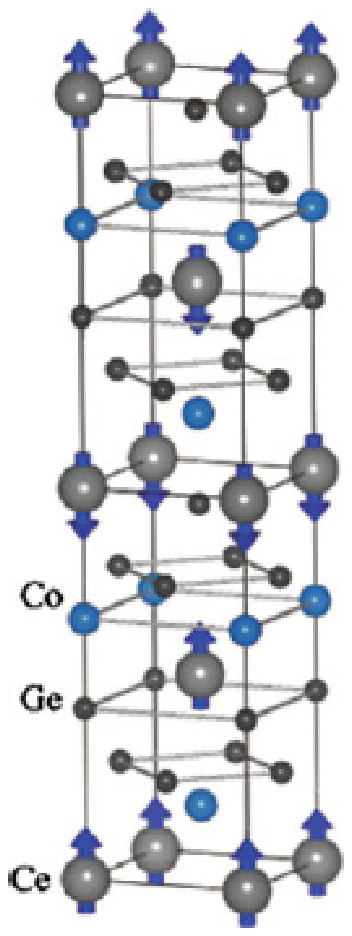}
\caption{Possible magnetic structure for $q_1$=(0,0,0.5)
sublattice in the ground state of CeCoGe$_3$ \cite{kaneko}. }
\label{fig:ccg_mag_st}
\end{figure}

The neutron-diffraction experiments revealed that the magnetic
structure of the ground state at ambient pressure consists of
two components with dominant $q_1$=(0,0,1/2) and subordinate
$q_2$=(0,0,3/4) \cite{kaneko}. Figure \ref{fig:ccg_mag_st}
shows a possible magnetic structure of the $q_1$ sublattice.
The magnetic moments are parallel to the $c$ axis and alternate
in the up-up-down-down sequence. The magnitude of the magnetic
moment in the sublattice $\mu_1$ is estimated to be
0.5(1)$\mu_B$. A more complete magnetic structure that includes
the $q_2$ sublattice is not clear at present.

%%%%%%%%%%%%%%%%% FIGURE 6 %%%%%%%%%%%%%%%%%%%
%%%%%%%%%%%%%%%%% FIGURE 6 %%%%%%%%%%%%%%%%%%%

The magnetic susceptibility of CeCoGe$_3$ does not show a peak
or a saturated behavior at low temperatures as observed in
CeRhSi$_3$ and CeIrSi$_3$. Specific heat measurements revealed
that the entropy gain reaches 68\% of $R\ln 2$ at $T_{{\rm
N}1}=21$ K and that 32\% of entropy loss is recovered at 38 K.
The magnetism of CeCoGe$_3$ is basically understood in terms of
localized 4{\em f} electron.

%-----------------------------------------------------------
\vspace{\baselineskip}
\noindent\textbf{CeIrGe$_3$}\\
The magnetic structure of CeIrGe$_3$ seems to be more
complicated. There are two successive magnetic transitions at
$T_{N1}$=8.7 K and $T_{N2}$=4.8 K. The former is
antiferromagnetic and the latter is unknown although
magnetization measurements indicate weak ferromagnetism with a
small moment below $T_{N2}$ (see Fig.~\ref{fig:mag_all}(d)). A
parasitic ferromagnetism due to the Dzyaloshinsky-Moriya
interaction caused by the broken space inversion symmetry is
discussed in Ref. \cite{kawai2}. $T_{N2}$ merges into $T_{N1}$
with the application of pressure (see Fig.~\ref{fig:PD}(d)).

%----------------------------------------------------------------
\subsubsection{Temperature-Pressure Phase Diagram}
\begin{figure}[tb]
%\sidecaption
	\includegraphics[width=0.9\textwidth]{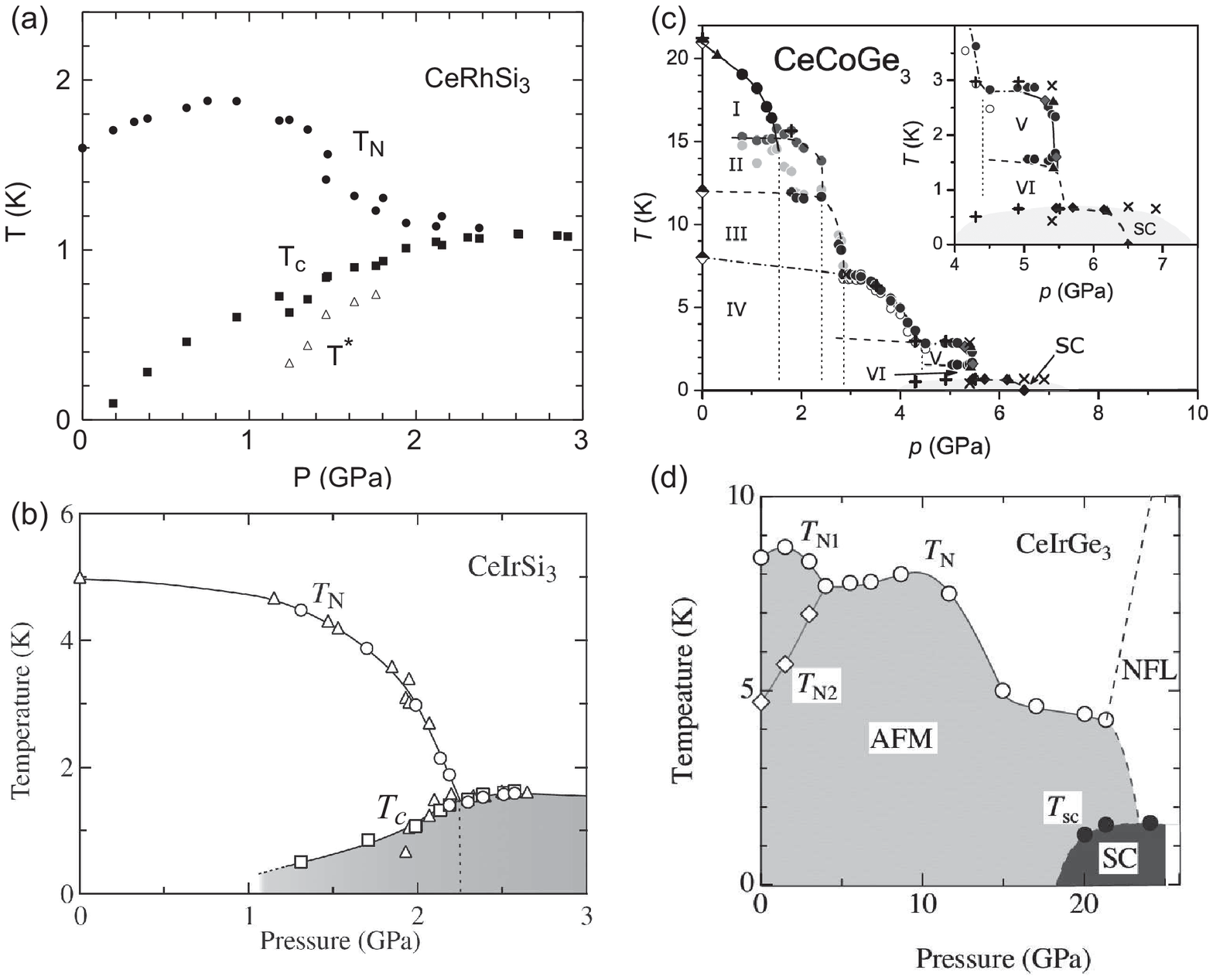}
\caption{Temperature-pressure phase diagrams of CeRhSi$_3$
\cite{kimura4}, CeIrSi$_3$ \cite{tateiwa2}, CeCoGe$_3$
\cite{knebel1} and CeIrGe$_3$ \cite{honda}. $T^*$ in panel (a)
denotes an anomaly observed in the resistivity and
magnetic-susceptibility measurements \cite{kimura1}. }
\label{fig:PD}
\end{figure}

All known Ce$TX_3$ superconductors need pressure to become
superconducting. The temperature-pressure ($T$-$P$) phase
diagrams of CeRhSi$_3$ \cite{kimura4}, CeIrSi$_3$
\cite{tateiwa2}, CeCoGe$_3$ \cite{settai} and CeIrGe$_3$
\cite{honda} are shown in Fig.~\ref{fig:PD}. In CeRhSi$_3$, the
N\'eel temperature $T_N$ initially increases and subsequently
decreases with applying pressure, while those of CeIrSi$_3$
decreases monotonically with pressure. The $T_N(P)$ of
CeCoGe$_3$ and CeIrGe$_3$ exhibits step-like decreases probably
relevant to successive magnetic phase transitions observed at
ambient pressure. The superconducting transitions of these four
compounds are observed at pressures at which the AFM order
still exists. We define here three characteristic pressures:
$P^*_1$ where $T_N=T_c$, $P^*_2$ where $T_N\rightarrow 0$ and
$P^*_3$ where the superconducting transition temperature $T_c$
reaches a maximum. The values of these pressures are summarized
in Table~\ref{t3}. $P^*_2$ of CeRhSi$_3$ is unclear because
$T_N$ does not decrease steeply near $P^*_1$.

%%%%%%%%%%%%%%%%% FIGURE 13 %%%%%%%%%%%%%%%%%%%
%%%%%%%%%%%%%%%%% FIGURE 13 %%%%%%%%%%%%%%%%%%%

\begin{table}[htb]
\caption{Normal and superconducting parameters of Ce$TX_3$
compounds. $\xi(0)$ is estimated from Eq.~(\ref{eq:orb}). Here,
we regard $H_{c2}(0)$ for $H\parallel c$ as $H_{orb}$; that is,
the paramagnetic pair-breaking effect is absent. $P^*_1$,
$P^*_2$ and $P^*_3$ are the pressures at which $T_N=T_c$,
$T_N\rightarrow 0$ and $T_c$ reaches a maximum, respectively.
$\vec q$, $\mathbf{m_Q}$ and $\mu_s$ denote magnetic
propagation vector, magnetic moment and value of ordered
moment, respectively, and $H_{c2}'\equiv
-dH_{c2}/dT|_{T=T_c}$.} 	
\begin{tabular}{p{3cm}p{2cm}p{2cm}p{2cm}p{2cm}}
\hline\noalign{\smallskip}
              & CeRhSi$_3$  & CeIrSi$_3$  & CeCoGe$_3$ & CeIrGe$_3$ \\
\noalign{\smallskip}\svhline\noalign{\smallskip}
		crystal structure                              & \multicolumn{4}{c}{tetragonal}  \\
		space group                                    & \multicolumn{4}{c}{$I4mm$}      \\
		$a$ [\AA ]                             & 4.237    & 4.252   & 4.320 & 4.401 \\
        $c$ [\AA ]                             & 9.785    & 9.715   & 9.835 & 10.024 \\
\noalign{\smallskip}\hline\noalign{\smallskip}
		$\gamma_n$ [mJ/mol$\cdot$K$^2$]              & 110      & 120     &  32 & 80 \\
        $m^*$ [$m_0$]                          & $4-19$   & N/A     &  N/A  & N/A \\
		$l$ [\AA ]                        & $2400-3400$ & N/A  &  N/A  & N/A \\
		$T_N$ [K]                        & 1.6      & 5.0     &  21/12/8 & 8.7/4.8 \\
		$\vec q$ &
$(\pm0.215,0,0.5)$ & N/A &
$(0,0,1/2)$ \par $(0,0,3/4)$ & N/A \\
        $\mathbf{m_Q}$ orientation &
        [001]  &  [001]  & [100] & [100]? \\
		$\mu_s$ [$\mu_B$/Ce] & 0.13 & N/A  & 0.5 & N/A \\
\noalign{\smallskip}\hline\noalign{\smallskip}
		$P^*_1$ (GPa)           & 2.4-2.5     & 2.25         & 5.5-5.6$^d$ & $>21$  \\
		$P^*_2$ (GPa)           & ?           & 2.50         & 5.5-5.7$^d$ & $\sim 24$  \\
		$P^*_3$ (GPa)           & 2.65        & 2.63         & 5.7$^d$ or 6.5$^e$ & $\geq 24$ \\
\noalign{\smallskip}\hline\noalign{\smallskip}
		$T_{c}$ @$P^*_3$ [K]                         & 1.09      & $1.56-1.59$ & 0.66$^d$ or 0.69$^e$ & $\geq 1.6$ \\
        $\Delta C/\gamma_nT_{c}$                      & N/A       & 5.7       &  N/A & N/A \\
		$H_{c2}(0)$ [T] ($H\parallel c$)       & $30\pm 2^a$ & $45\pm 10^c$ & $22\pm 8^{f,g}$ & $27\pm 10$\\
		$H_{c2}(0)$ [T] ($H\perp c$)                   & $7.5^b$   & $9.5^c$   &  N/A  & N/A \\
		$\xi(0)$ [\AA ] ($H\parallel c$)                & $\sim 33$ & $\sim 27$ & $\sim 39$  & $\sim 35$ \\
		$H_{c2}'$ [T/K] ($H\parallel c$)  & $23^b$    & $17^c$    & $20^d$  & $16^h$      \\
		$H_{c2}'$ [T/K] ($H\perp c$)      & $27^b$    & $14.5^c$  & N/A     & N/A   \\
\noalign{\smallskip}\hline\noalign{\smallskip}		
\end{tabular}\\
$^a$ at 2.85 GPa\\
$^b$ at 2.6 GPa\\
$^c$ at 2.65 GPa\\
$^d$ determined from heat-capacity measurements\cite{knebel1}.\\
$^e$ determined from electrical-resistivity measurements\cite{kawai2}.\\
$^f$ at 6.5 GPa \\
$^g$ estimated from Fig.~\ref{fig:ccg_hc2}.\\
$^h$ at 24 GPa	\label{t3}
\end{table}

The resistivity drop at the superconducting transition of
CeRhSi$_3$, CeIrSi$_3$ and probably CeCoSi$_3$ has its sharpest
form at $P^*_3$. The resistivity drop in the AFM phase,
especially far below $P^*_3$, is very broad. In this pressure
region, the drop width depends on the applied current
 \cite{kimura1,tateiwa2}. These observations imply that
superconductivity is inhomogeneous or fluctuating in the
antiferromagnetic state and is optimum at $P^*_3$. Such a
phenomenon is found in centrosymmetric HF superconductors as
well and, thus, seems to be realized irrespective of the
presence or absence of inversion symmetry.

\begin{figure}[tb]
%\sidecaption
	\includegraphics[width=0.9\textwidth]{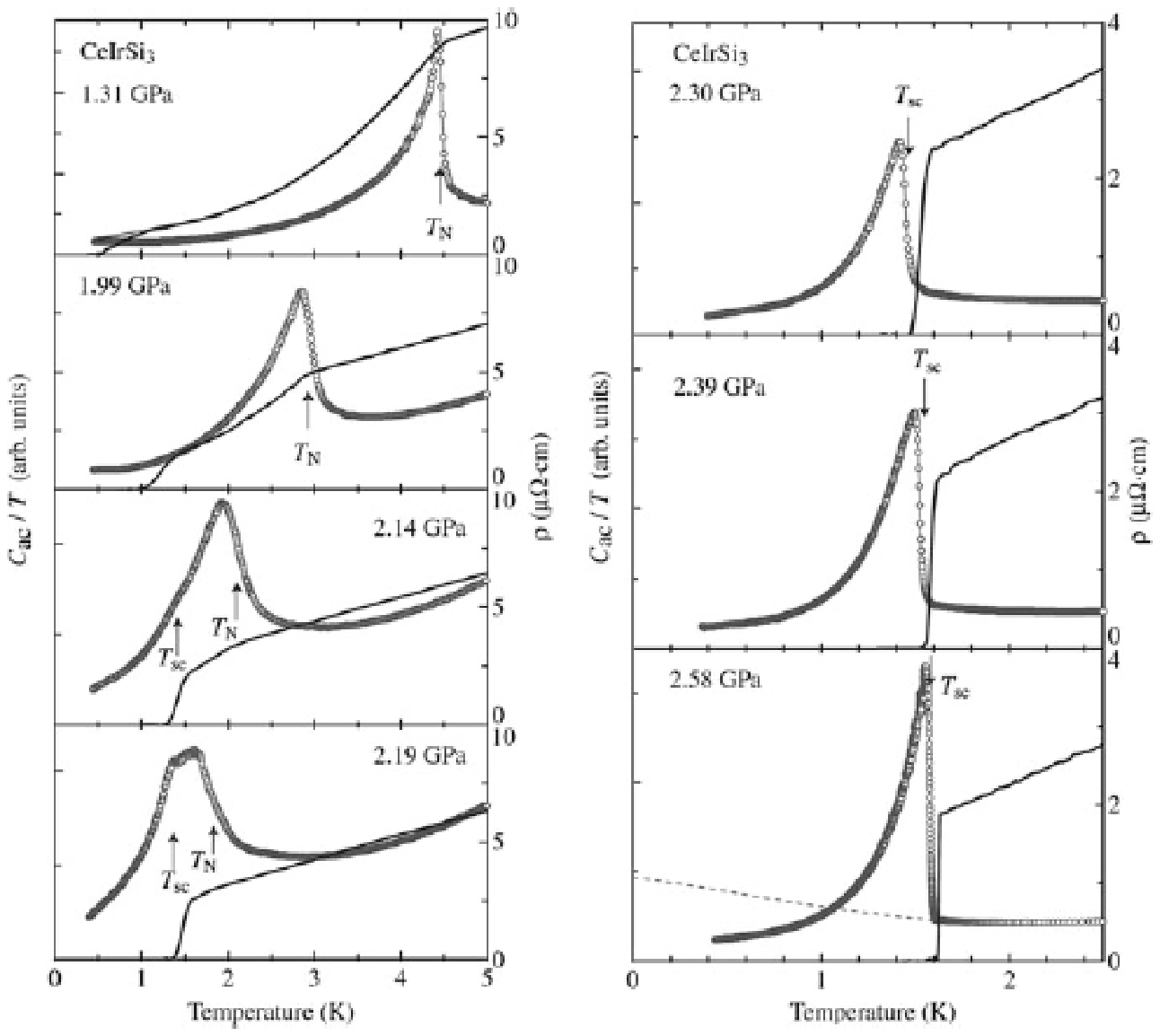}
\caption{Temperature dependence of the ac heat capacity
$C_{ac}$ (circles, left side) and the electrical resistivity
$\rho$ (lines, right side) at several pressures in CeIrSi$_3$.
The dotted line in the panel at 2.58 GPa indicates the entropy
balance below $T_c$ \cite{tateiwa2}. } \label{fig:cis_C_super}
\end{figure}

A similar evolution of superconductivity in the AFM state is
seen by the heat capacity $C$ (Fig.~\ref{fig:cis_C_super}). The
heat capacity jump ($\Delta C$) at the superconducting
transition below $P^*_1$ is small and broad, while it becomes
sharper with increasing pressure and is strongly enhanced above
$P^*_1$ \cite{tateiwa2}. In CeIrSi$_3$, as shown in
Fig.~\ref{fig:cis_C_super}, $\Delta C/C_n$, where $C_n$ is the
normal state value just above $T_c$, reaches $5.7\pm0.1$ at
2.58 GPa, which is much larger than the 1.43 value expected
from the weak-coupling BCS model and is probably the highest
value among all known superconductors. On the other hand, the
jump of the heat capacity associated with antiferromagnetism
diminishes when approaching $P^*_1$, and is no longer observed
above $P^*_1$. Apparently, the entropy gain of the AFM
transition is transferred to the superconducting one.
%%%%%%%%%%%%%%%%% FIGURE 13 %%%%%%%%%%%%%%%%%%%
%%%%%%%%%%%%%%%%% FIGURE 13 %%%%%%%%%%%%%%%%%%%

%------------------------------------------------------------------
\subsubsection{Quantum Criticality and Non-Fermi Liquid}
\begin{figure}[tb]
%\sidecaption
	\includegraphics[width=0.9\textwidth]{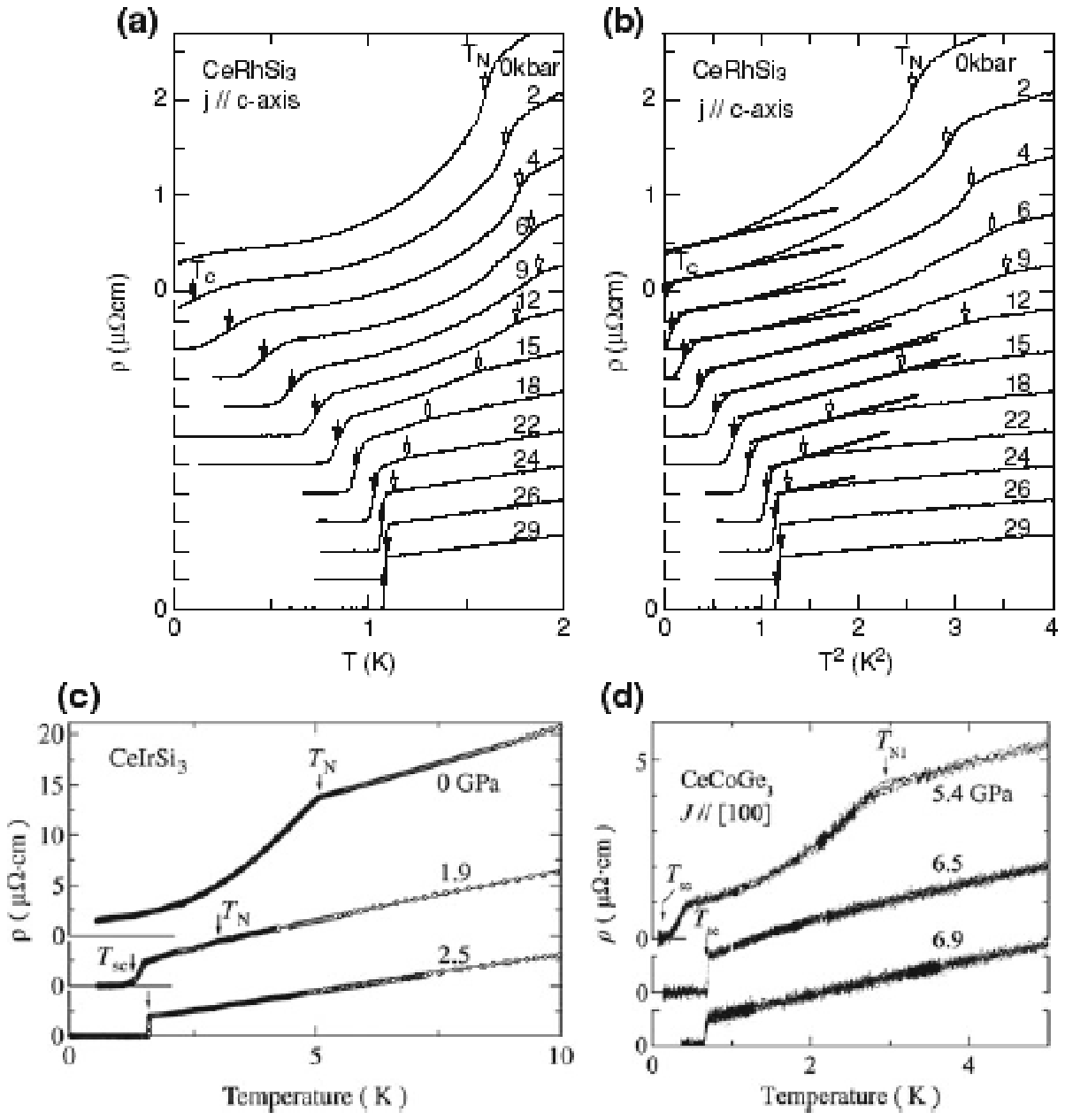}
\caption{Temperature dependence of the resistivity of (a)
CeRhSi$_3$ \cite{kimura4}, (c) CeIrSi$_3$ \cite{sugitani} and
(d) CeCoGe$_3$ \cite{kawai2}. (b) Resistivity against $T^2$ in
CeRhSi$_3$. } \label{fig:resis}
\end{figure}

Superconductivity in $f$-electron materials often appears in
the vicinity of a quantum critical point (QCP) at which the
magnetic ordering temperature is reduced to zero by a
nonthermal control parameter such as pressure, magnetic field,
or chemical substitution. Often, a QCP accompanies the
emergence of a non-Fermi liquid in which the temperature
dependence of some physical properties obeys different behavior
from that expected in the Fermi-liquid theory. As shown in
Fig.~\ref{fig:resis}, the electrical resistivity changes from
the Fermi-liquid prediction $\rho(T)=\rho_0+AT^2$ below $P^*_1$
to $\rho(T)=\rho_0+A^{\prime}T$ above $P^*_1$ in CeRhSi$_3$ and
CeIrSi$_3$ \cite{sugitani,kimura4}. A similar crossover occurs
in CeCoGe$_3$ at 6.9 GPa \cite{kawai2}. The $T$-linear
dependence of the resistivity agrees with the prediction by the
2D spin fluctuation theory as indicated in Table~\ref{t2}. On
the other hand, in CeIrSi$_3$ $1/T_1$-NMR measurements in the
normal state reveal a $\sqrt{T}$ dependence at 2.7 GPa for
$H\perp c$. This supports a prediction for 3D AFM fluctuations
in this system: $1/T_1\propto T\sqrt{\chi_Q(T)}\propto
T/\sqrt{(T+\theta)}$
 \cite{mukuda}. Here, $\chi_Q(T)$, the staggered susceptibility
with the AFM propagation vector $\vec{Q}$, follows the
Curie-Weiss law and $\theta$ measures the deviation from the
QCP.
%%%%%%%%%%%%%%%%% FIGURE 14 %%%%%%%%%%%%%%%%%%%
%%%%%%%%%%%%%%%%% FIGURE 14 %%%%%%%%%%%%%%%%%%%
%%%%%%%%%%%%%%%%% FIGURE 14-2 %%%%%%%%%%%%%%%%%%%
\begin{figure}[tb]
%\sidecaption
	
\includegraphics[width=0.5\textwidth]{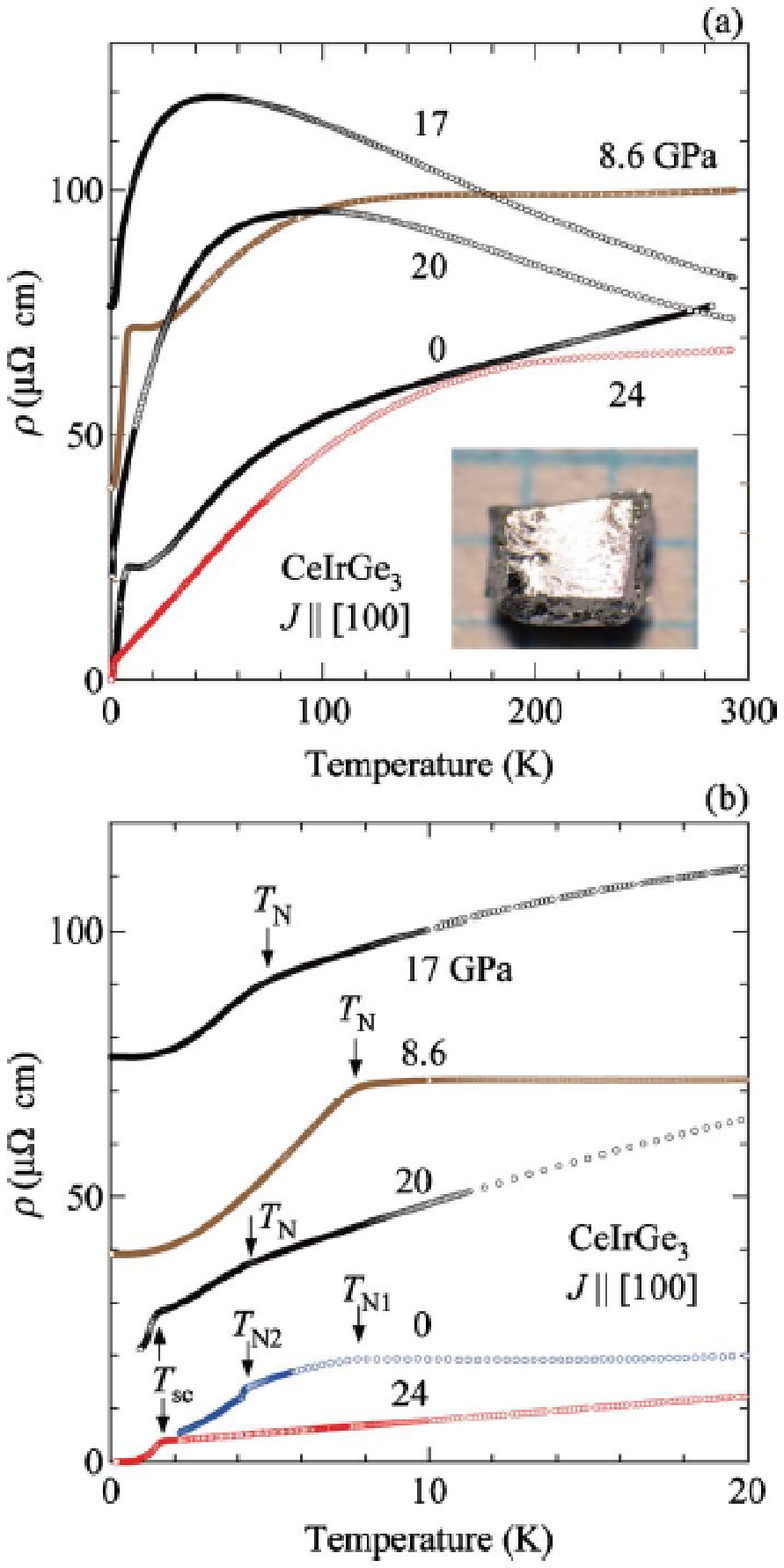}
\caption{Temperature dependence of the resistivity of
CeIrGe$_3$ at several pressures up to 24 GPa (a) below room
temperature and (b) below 20 K \cite{honda}. }
\label{fig:resis_cig}
\end{figure}
%%%%%%%%%%%%%%%%% FIGURE 14-2 %%%%%%%%%%%%%%%%%%%
The specific heat divided by temperature $C/T$ of CeIrSi$_3$ is
estimated to be enhanced almost linearly with decreasing
temperature from consideration of the entropy balance
\cite{tateiwa2}. The enhancement of $C/T$ is consistent with
the spin fluctuation theory, even though from theory we cannot determine
the proper dimensionality. The contradiction in the
temperature responses of $\rho$ and $1/T_1$ remains unsettled.

In general, it is unclear whether or not a QCP truly exists in
Ce$TX_3$, since the AFM transitions above $P^*_1$ are prevented
or masked by superconductivity. For example, the specific heat
jump associated with the AFM transition seems to disappear
suddenly just above $P^*_1$ in CeIrSi$_3$ as mentioned above.
This suggests that the QCP exists neither at $P^*_2$ nor at
higher pressures. However, this does not necessarily indicate
that magnetic fluctuations vanish at these pressures. The fact
that the maximum $T_c$ and the sharpest resistivity drop take
place at $P^*_3$ implies that magnetic fluctuations survive
even above $P^*_1$ and rather develop toward $P^*_3$. Assuming
that magnetic fluctuations stabilize the superconducting phase
and become stronger at the QCP, $P^*_3$ should be regarded as a
{\em virtual} QCP.

In the vicinity of the QCP some pressure-induced HF
superconductors display a strong enhancement of the effective
mass of the conduction electrons; namely, via the $\gamma_n$
value and the coefficient $A$ of the $T^2$ term of the
electrical resistivity \cite{fisher,araki,knebel}. In the cases
of CeRhSi$_3$ and CeIrSi$_3$, however, such an enhancement is
less obvious. The coefficient $A$ in CeRhSi$_3$ is almost
constant up to $P^*_1$ \cite{kimura4} and the $\gamma_n$ value
in CeIrSi$_3$ is suggested to be unchanged up to $P^*_3$
\cite{tateiwa2}. On the other hand, the coefficient $A$ in
CeCoGe$_3$ is strongly enhanced from 0.011
$\mu\Omega\cdot$cm/K$^2$ at ambient pressure to 0.357
$\mu\Omega\cdot$cm/K$^2$ at 5.4 GPa. In CeIrGe$_3$ $\rho(T)$
exhibits a complicated pressure dependence (as shown in
Fig.~\ref{fig:resis_cig}) that makes unclear how $A$ varies
with pressure. The drastic changes at certain pressures between
8.6 GPa and 17 GPa may be associated with the step-like
decrease of $T_N(P)$ around 13 GPa as seen in
Fig.~\ref{fig:PD}(d). The residual resistivity becomes maximum
at a pressure near 17 GPa. Since such an enhancement of
residual resistivity can be a signature of critical valence
fluctuations \cite{miyake}, a valence transition or crossover
may take place around 13 GPa in CeIrGe$_3$.

%%%%%%%%%%%%%%%%% TABLE II %%%%%%%%%%%%%%%%%%%%
\begin{table}[tb]
\caption{ Theoretically predicted quantum critical behavior for
3D and 2D antiferromagnetic fluctuations and corresponding
temperature dependence on Ce$TX_3$ compounds. } \label{t2}
\begin{tabular}{p{2cm}p{2.4cm}p{2.4cm}p{2.4cm}}%{cccc}
\hline\noalign{\smallskip}
           & $\rho$           & $1/T_1$             & $C/T$            \\
\noalign{\smallskip}\svhline\noalign{\smallskip}
3D AFM     & $T^{3/2}$        & $T/\sqrt{T+\theta}$ & const.$-T^{1/2}$ \\
2D AFM     & $T$              & $T/(T+\theta)$      & $-\ln T$
\\ \noalign{\smallskip}
CeRhSi$_3$ & $T$ \cite{kimura1}&                     &                  \\
CeIrSi$_3$ & $T$ \cite{sugitani}& $T/\sqrt{T+\theta}$ \cite{mukuda} & $T$ ? \cite{tateiwa2}            \\
CeCoGe$_3$ & $T$ \cite{kawai2} &                     &                  \\
CeIrGe$_3$ & $T$ \cite{honda}  &                     &                  \\
\noalign{\smallskip}\hline\noalign{\smallskip}
\end{tabular}
\end{table}
%%%%%%%%%%%%%%%%% TABLE II %%%%%%%%%%%%%%%%%%%%

%
%%%%%%%%%%%%%%%%%%%%%%%%%%%%%%%%%%%%%%%%%%%%%%%%%%%%%%%%%%%%%%%%%%%%%%%%%%%%%%%
\subsection{Superconducting State}
%
%------------------------------------------------------------------
\subsubsection{Anisotropic Upper Critical Field : Two Limiting Fields}
Thus far, the most interesting phenomenon in the Ce$TX_3$
superconductors is the extremely high anisotropy in the upper
critical magnetic field $H_{c2}$, with stunningly high values
for fields along the $c$ axis. In CeRhSi$_3$ and CeIrSi$_3$,
for example, $H_{c2}$ exceeds 30 T along the $c$ axis, whereas
falls below 10 T along the plane. Considering that in these
materials $T_c$ is of the order of 1 K, this leads to very high
$H_{c2}/T_c$ ratios not known previously for any
\emph{centrosymmetric} superconductors (except for
field-induced superconductors like URhGe \cite{levy} and
organic superconductors \cite{uji}). The strong field
anisotropy and the extremely high $H_{c2}$s are thought to be
understood in terms of the anisotropy of the paramagnetic
pair-breaking effect characteristic of non-centrosymmetric
(Rashba-type) superconductors.

There are two pair-breaking mechanisms for Cooper pairs under
magnetic fields: the paramagnetic and the orbital. The former
mechanism is attributed to the spin polarization due to the
Zeeman effect, which competes with the antiparallel-spin
formation of the Cooper pair in spin-singlet superconductors.
The influence of the paramagnetic effect depends on the
symmetry of the Cooper pairs, as discussed later. On the other
hand, the orbital effect is ascribed to the orbital motion in a
magnetic field. The influence of this effect is thought to be
independent of the pairing symmetry. The magnitude of
$H_{c2}(0)$ is consequently restricted by both the paramagnetic
(Pauli-Clogston-Chandrasekhar) limiting field $H_P$ and the
orbital limiting field $H_{orb}$ \cite{hake}. Hereafter, we
call $H_P$ the Pauli limiting field for simplicity.

\begin{figure}[tb]
%\sidecaption
	
\includegraphics[width=0.9\textwidth]{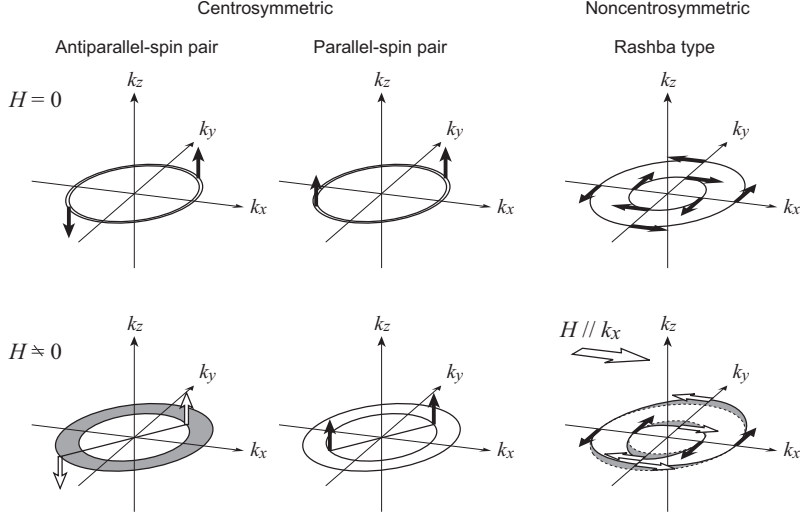}
\caption{Two-dimensional Fermi surfaces for centrosymmetric and
non-centrosymmetric superconductors at zero and finite fields.
The centrosymmetric superconductors can be classified into
parallel- and antiparallel-spin pairs. In the
non-centrosymmetric superconductor, only the Rashba type is
displayed. The hollow arrows indicate a pairing no longer
allowed. In the antiparallel-spin pair including the
conventional singlet pair, the pair breaking occurs on the
whole Fermi surface under magnetic fields. On the other hand,
in the parallel-spin pair, all the pairing persist. In the
Rashba-type superconductor, pair breaking occurs only for the
pairs parallel to the applied magnetic field. When the magnetic
field is applied to $k_z$, pair breaking does not occur. }
\label{fig:anisopauli}
\end{figure}

%-----------------------------------------------------------
\vspace{\baselineskip}
\noindent\textbf{Pauli-Clogston-Chandrasekhar limit}\\
The paramagnetic effect in spin-singlet and spin-triplet
pairing symmetries is illustrated schematically in
Fig.~\ref{fig:anisopauli}. In centrosymmetric superconductors,
the up-spin and down-spin bands of the conduction electrons are
degenerate in zero field. The corresponding Fermi surfaces are
also degenerate. The presence of a magnetic field inflates and
deflates the down-spin and up-spin Fermi surfaces,
respectively, due to the Zeeman effect. When the Cooper pair
consists of antiparallel spins, e.g.~the conventional singlet
pair, the paramagnetic pair breaking occurs on the whole Fermi
surface. When the Cooper pair comprises parallel spins, namely
the triplet pair, the paramagnetic pair breaking does not
occur, therefore the Pauli limit is absent. Since the spin
direction of the Cooper pair is always aligned to the magnetic
field, both cases are independent of the field direction unless
the coupling between the orbital and spin parts of the pairing
function is present.
%%%%%%%%%%%%%%%%% FIGURE 17 %%%%%%%%%%%%%%%%%%%
%%%%%%%%%%%%%%%%% FIGURE 17 %%%%%%%%%%%%%%%%%%%

In non-centrosymmetric superconductors, the up-spin and
down-spin Fermi surfaces are not degenerate even in zero field.
In the Rashba-type (tetragonal) superconductors, the spins are
perpendicularly aligned with the momenta in the $k_z$ plane
(Fig.~\ref{fig:anisopauli}) by the spin-orbit coupling,
yielding a momentum-dependent effective magnetic field. The
presence of an applied magnetic field along the $k_x$ direction
inflates and deflates the Fermi surfaces only along the $k_y$
direction, since the spin component perpendicular to the $k_x$
direction is not affected by this field. The paramagnetic
pair-breaking effect is thus partial for field along the $k_z$
plane. On the other hand, when the magnetic field is applied
along the $k_z$ axis, the paramagnetic pair-breaking effect is
absent because all the spins aligned with the $k_z$ plane are
perpendicular to the field direction. The strongly anisotropic
$H_{c2}(T)$, with high $H_{c2}(0)$ for $H\parallel c$, realized
in Ce$TX_3$ compounds is attributed to the anisotropic spin
susceptibility expected to appear in non-centrosymmetric
superconductors as discussed above in Spin State,
Sect.~\ref{pairingsys}.

\begin{figure}[tb]
%\sidecaption
	\includegraphics[width=0.8\textwidth]{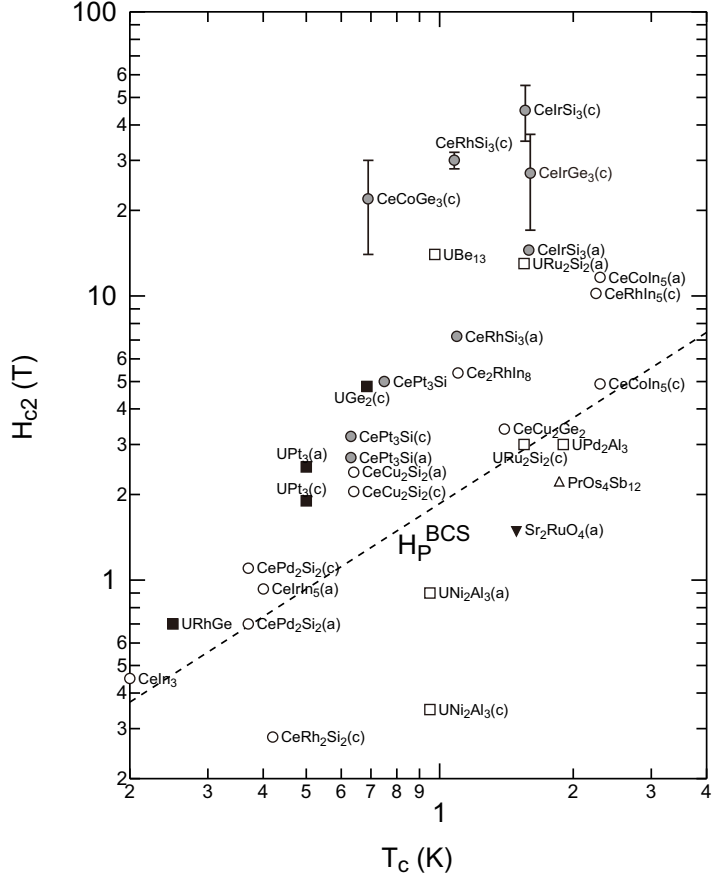}
\caption{$H_{c2}(0)$ versus $T_c$ for heavy-fermion and some
other well-known superconductors. The broken line represents
$H_P^{\rm BCS}=1.86T_c$. The circles, squares and upper and
lower triangles indicate cerium, uranium, praseodymium and
other compounds, respectively. The solid and hollow symbols tag
possible triplet and singlet (or unclear) superconductors,
respectively. The gray marks label non-centrosymmetric
superconductors. The letter $a$ or $c$ in parentheses denotes
the applied field direction and no letter indicates that the
result was obtained for a polycrystal. } \label{fig:pauli}
\end{figure}

Figure~\ref{fig:pauli} shows $H_{c2}(0)$ versus $T_c$ for
Ce$TX_3$ compounds and some well-known HF superconductors. The
dashed line indicates $H_P^{\rm BCS}$ (see Eq.~\ref{eq:hpbcs}).
The $H_{c2}(0)$ of the U-based superconductors UGe$_2$, URhGe
and UPt$_3$ exceeds $H_{P}^{\rm BCS}$. These superconductors
are thought to form a parallel-spin pairing free from the
paramagnetic pair-breaking effect. In some spin-singlet
superconductors $H_{c2}(0)$ is located above the $H_P^{\rm
BCS}$ line, which seems a contradiction. Two possibilities have
been proposed: one is a reduction of the $g$-factor and the
other is an enhancement of $\Delta_0$ due to a strong-coupling
effect. $H_{c2}(0)$s of Ce$TX_3$ compounds do not only exceed
the $H_P^{\rm BCS}$ line, but also surpass those of all other
materials. To understand the high $H_{c2}$ of the Ce$TX_3$
compounds, it is necessary to consider the orbital
pair-breaking effect that we discuss next.
%%%%%%%%%%%%%%%%% FIGURE 18 %%%%%%%%%%%%%%%%%%%
%%%%%%%%%%%%%%%%% FIGURE 18 %%%%%%%%%%%%%%%%%%%

%-------------------------------------------------------------------------
\vspace{\baselineskip}
\noindent\textbf{Orbital limit}\\
As we discussed above, the orbital limiting field can be
estimated from Eq.~(\ref{eq:initial}). The value $h(0)$ depends
on both the ratio $\xi(0)/l$ and the strong-coupling parameter
$\lambda$ ($l$: mean free path). In the weak-coupling limit
($\lambda=0$), namely the BCS model, $h(0)=0.727$ for
$(\xi(0)/l)\rightarrow 0$ (clean limit) and $h(0)=0.693$ for
$(\xi(0)/l) \rightarrow\infty$ (dirty limit). Most HF
superconductors satisfy the clean limit. In the strong-coupling
limit ($\lambda\to \infty$), $h(0)$ approaches 1.57 for clean
superconductors \cite{bulaevskii} and increases with $\lambda$
for dirty superconductors. It is noted that the $\lambda$
dependence of $h$ is usually derived on the basis of the
conventional electron-phonon model, but that in HF systems
 a corresponding electron-magnon approach should be employed since
 it is generally believed that the attractive interaction leading to
 Cooper pairs arises from coupling to spin excitations.

\begin{figure}[tb]
%\sidecaption
	\includegraphics[width=0.8\textwidth]{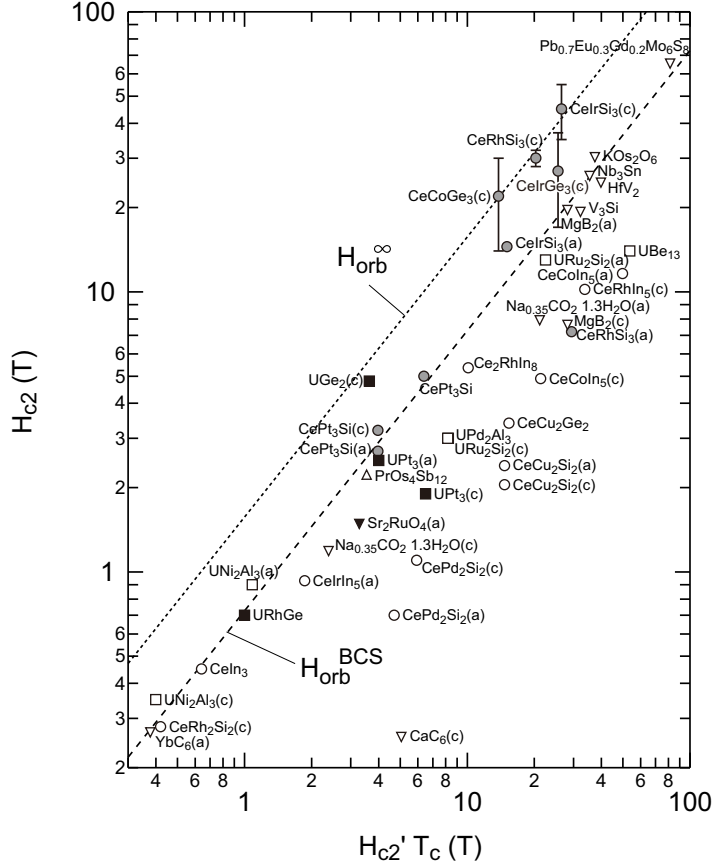}
\caption{$H_{c2}(0)$ versus $H_{c2}^{\prime}T_c$  for
heavy-fermion and other well-known superconductors. The broken
and dotted lines represent $H_{orb}^{\rm
BCS}=0.727H_{c2}^{\prime}T_c$ and
$H_{orb}^{\infty}=1.57H_{c2}^{\prime}T_c$, respectively. The
circles, squares and upper and lower triangles indicate cerium,
uranium, praseodymium and other compounds, respectively. The
solid and hollow symbols tag possible triplet and singlet (or
unclear) superconductors, respectively. The gray marks label
non-centrosymmetric superconductors. The letter $a$ or $c$ in
parentheses denotes the applied field direction and no letter
indicates that the result was obtained for a polycrystal. }
\label{fig:orbital}
\end{figure}

The $H_{c2}(0)$s in Fig.~\ref{fig:pauli} are plotted against
$H_{c2}^{\prime}T_c$ in Fig.~\ref{fig:orbital}. The dashed and
dotted lines indicate the orbital limiting fields at
$\lambda=0$ (weak-coupling limit) and
$\lambda\rightarrow\infty$ (strong-coupling limit) for clean
superconductors; that is, $H_{orb}^{\rm
BCS}=0.727H_{c2}^{\prime}T_c$ and
$H_{orb}^{\infty}=1.57H_{c2}^{\prime}T_c$, respectively. Some
parallel-spin, namely triplet, superconductors like UPt$_3$ and
URhGe are located just below $H_{orb}^{\rm BCS}$. The compounds
located far below the $H_P^{\rm BCS}$ line in
Fig.~\ref{fig:pauli}, e.g.~UNi$_2$Al$_3$ and CeRh$_2$Si$_2$,
are seen in the vicinity of the $H_{orb}^{\rm BCS}$ line in
Fig.~\ref{fig:orbital}. The $H_{c2}$s of these compounds may be
mainly restricted by the orbital limit rather than by the Pauli
limit. CePt$_3$Si is also located near the $H_{orb}^{\rm BCS}$
line. Considering that  $H_{c2}$ is almost isotropic in
CePt$_3$Si \cite{yasuda}, it may be also mainly constrained by
the orbital limit rather than by the paramagnetic one.

The $H_{c2}$s of the Ce$TX_3$ compounds for fields along the
$c$ axis well exceed the $H_{orb}^{\rm BCS}$ line. They seem to
be close to the strong-coupling limit $H_{orb}^{\infty}$.
Although the high $H_{c2}/(H_{c2}^{\prime}T_c)$ is a common
feature in the Ce$TX_3$ HF superconductors, it is not obvious
that such a result can be associated with their
non-centrosymmetric crystal structures. Note that UGe$_2$ also
seems to be above the $H_{orb}^{\rm BCS}$ line, but this is due
to the jump of the $H_{c2}$ curve attributed to the
metamagnetic transition \cite{huxley, sheikin}. Therefore, this
plot does not necessarily indicate that the intrinsic
$H_{c2}(0)$ of UGe$_2$ exceeds the $H_{orb}^{\rm BCS}$ line.
%%%%%%%%%%%%%%%%% FIGURE 19 %%%%%%%%%%%%%%%%%%%
%%%%%%%%%%%%%%%%% FIGURE 19 %%%%%%%%%%%%%%%%%%%

%---------------------------------------------------------------
\subsubsection{Upper Critical Field for $c$ axis}
\begin{figure}[]
%\sidecaption
	\includegraphics[width=0.65\textwidth]{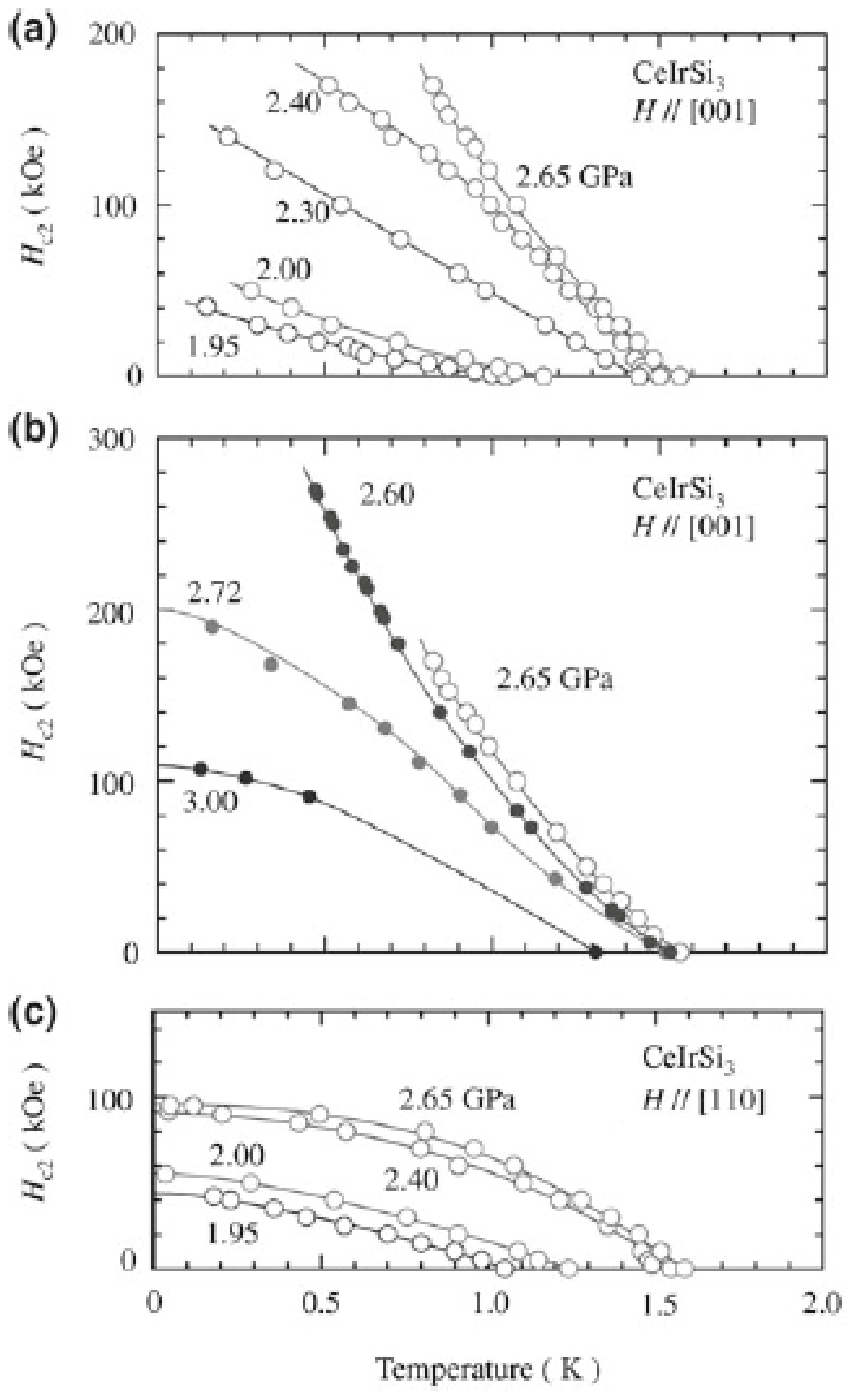}
\caption{$H_{c2}$-$T$ phase diagrams of CeIrSi$_3$ for
$H\parallel c$ at $P\le P^*_3\approx 2.65$ GPa (a), at $P\ge
P^*_3$ (b), and for $H\perp c$ at $P\le P^*_3$ \cite{settai1}.
} \label{fig:cis_hc2}
\end{figure}

In addition to the high values of $H_{c2}(0)$, the upward shape
of the temperature dependence of $H_{c2}(T)$ for certain
pressures seems to be also a characteristic of the Ce$TX_3$
superconductors \cite{kawai2,settai1,kimura4}. They mostly keep
a positive curvature ($d^2H_{c2}/dT^2>0$) down to relatively
low temperatures; for example, in CeIrSi$_3$  down to $T\approx
0.25T_c$ at 2.6 GPa (Fig. \ref{fig:cis_hc2}). Interestingly,
the curve shapes of CeRhSi$_3$ and CeIrSi$_3$ vary with
pressure in a complex manner. As for CeIrSi$_3$, a positive
curvature is seen up to 2 GPa that gradually changes to a
negative curvature at 2.4 GPa to a quasi-linear shape at 2.3
GPa. At 2.6 GPa and 2.65 GPa, a positive curvature is recovered
that turns to a negative one at higher pressures. Similar
behavior is seen in CeRhSi$_3$. Below and above 2.6 GPa, the
curvatures of $H_{c2}(T)$ are positive. At 2.6 GPa, a
quasi-linear change of $H_{c2}(T)$ is observed. Because of the
lack of sufficient pressure data, it is not clear whether such
a phenomenon is realized in CeCoGe$_3$ and CeIrGe$_3$.

\begin{figure}[t]
%\sidecaption
	
\includegraphics[width=0.6\textwidth]{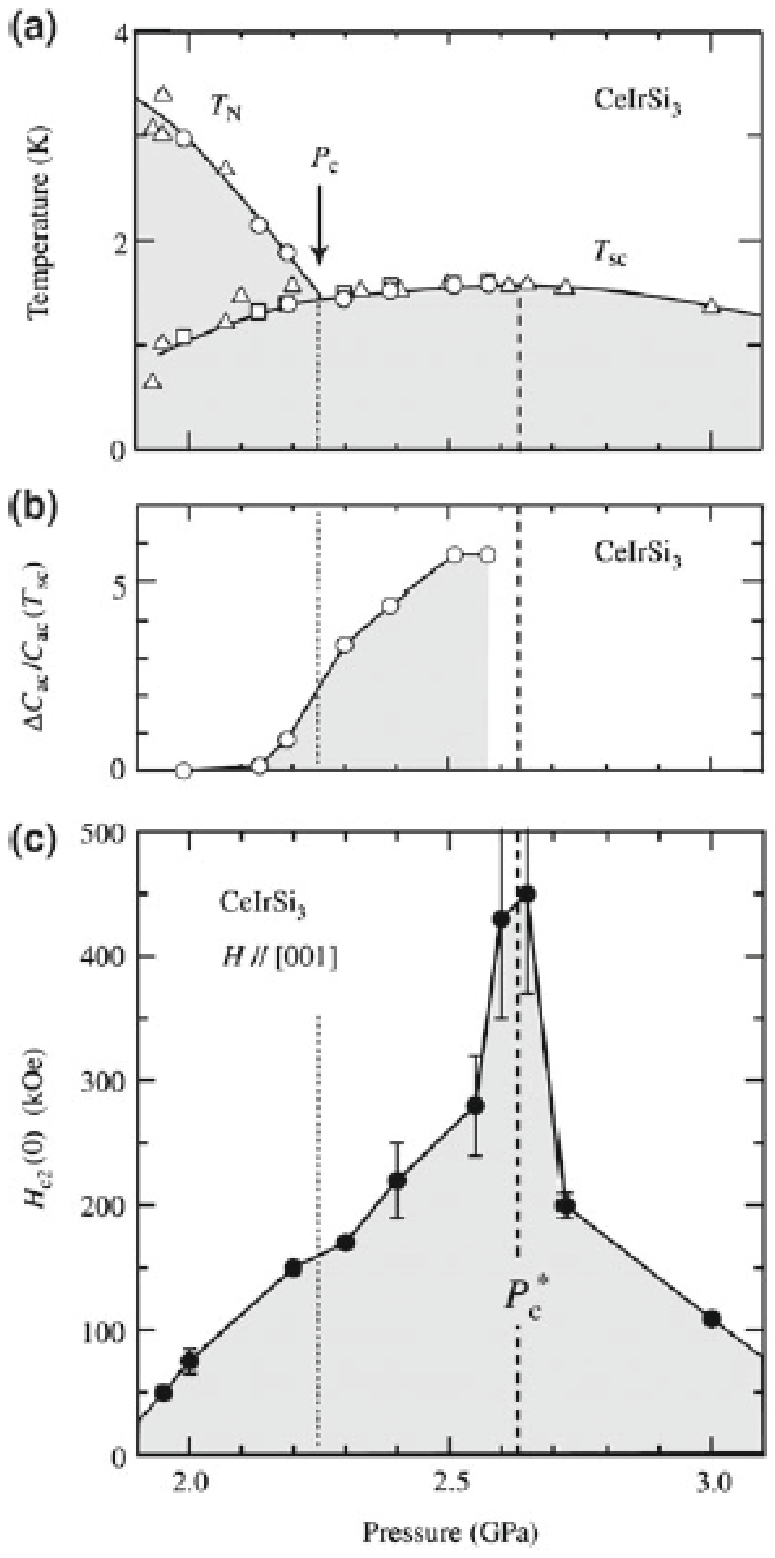}
\caption{Pressure dependence of (a) $T_N$ and $T_c$, (b)
specific heat jump $\Delta C/C(T_c)$ and (c) $H_{c2}(0)$ for
$H\parallel c$ in CeIrSi$_3$ \cite{settai1}. }
\label{fig:cis_pd_super}
\end{figure}

\begin{figure}[]
%\sidecaption
	\includegraphics[width=0.6\textwidth]{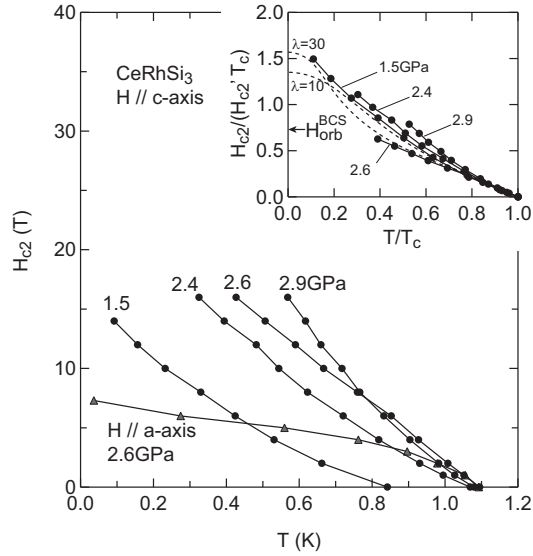}
\caption{$H_{c2}$-$T$ phase diagrams of CeRhSi$_3$ for
$H\parallel c$ at several pressures and for $H\parallel a$ at
$P^*_3\approx 2.6$ GPa. Inset: $H_{c2}(T)$ curves for
$H\parallel c$ normalized by the initial slope. The arrow
indicates the orbital limit $H_{orb}^{\rm BCS}$. The dashed
curves are theoretical predictions based on the strong-coupling
model using the coupling strength parameter $\lambda=10$ and 30
 \cite{kimura3}. } \label{fig:crs_hc2}
\end{figure}

\begin{figure}[]
%\sidecaption
	\includegraphics[width=0.6\textwidth]{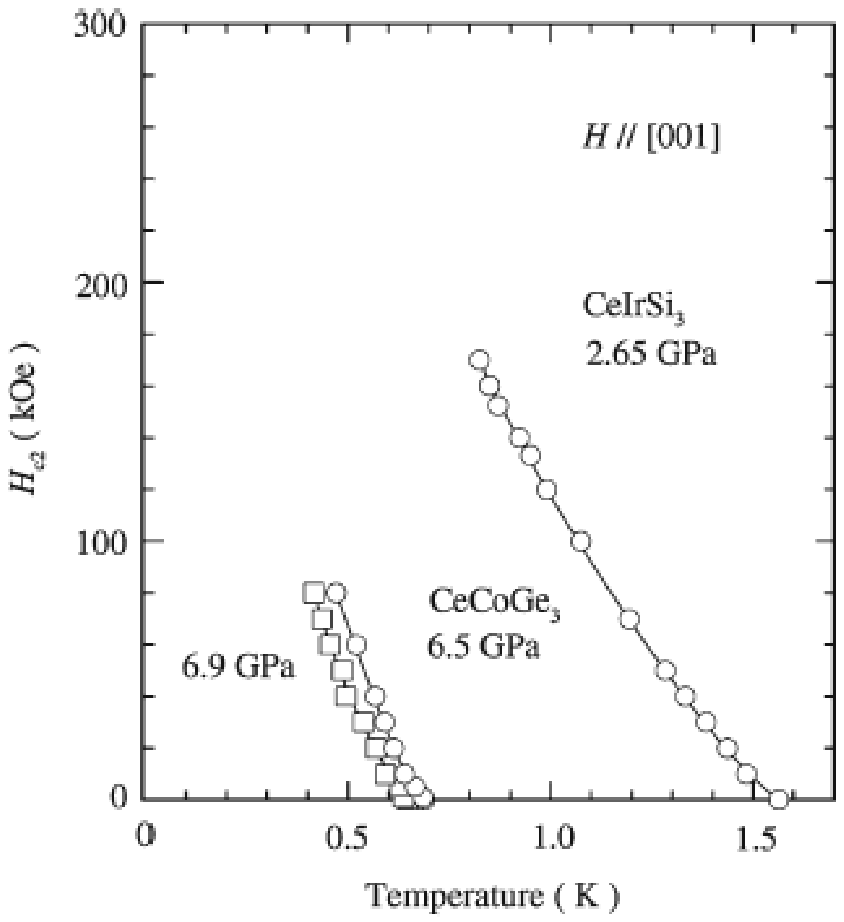}
\caption{$H_{c2}$-$T$ phase diagram of CeCoGe$_3$ for
$H\parallel c$ at 6.5 GPa and 6.9 GPa. The phase diagram of
CeIrSi$_3$ is also displayed \cite{kawai2}. }
\label{fig:ccg_hc2}
\end{figure}

In order to understand the superconducting phase diagram of the
Ce$TX_3$ compounds the pressure dependence of $H_{c2}(0)$ will
be very helpful. Fig.~\ref{fig:cis_pd_super}(c) shows
$H_{c2}(0)$ versus pressure in CeIrSi$_3$ \cite{settai1}.
%%%%%%%%%%%%%%%%% FIGURE 23 %%%%%%%%%%%%%%%%%%%
%%%%%%%%%%%%%%%%% FIGURE 23 %%%%%%%%%%%%%%%%%%%
$H_{c2}(0)$ increases with pressure and tends to diverge close
to 2.65 GPa ($\approx P^*_3$). Above 2.65 GPa it falls steeply
to half or less of the value of the maximum $H_{c2}(0)$. It is
pointed out that such an acute enhancement of $H_{c2}(0)$ can
be interpreted as an electronic instability arising at $P^*_3$
 \cite{settai1}. This instability can cause a mass
enhancement of the conduction electrons, giving rise to a
reduction in the superconducting coherence length $\xi(0)$. At
a first glance, a mass enhancement at $P^*_3$ is consistent
with the pressure dependence of the initial slope of the
superconducting $H$-$T$ phase diagram,
$H_{c2}^{\prime}=-dH_{c2}/dT|_{T=T_c}$. From
Eqs.~(\ref{eq:orb}) and (\ref{eq:initial}), we can derive
\begin{equation}
H_{c2}^{\prime}T_c\sim H_{orb}\sim \xi^{-2}(0)\sim
(\Delta_0m^*)^2 \,.
\end{equation}
\noindent Here, we use $v_{\rm F}=\hbar k_{\rm F}/m^*$. As
shown in Figs.~\ref{fig:cis_hc2} - \ref{fig:ccg_hc2},
$H_{c2}^{\prime}$s of CeIrSi$_3$ and CeRhSi$_3$ increase with
increasing pressure and become maximum at about $P^*_3$.
%%%%%%%%%%%%%%%%% FIGURE 21 %%%%%%%%%%%%%%%%%%%
%%%%%%%%%%%%%%%%% FIGURE 21 %%%%%%%%%%%%%%%%%%%
%%%%%%%%%%%%%%%%% FIGURE 20 %%%%%%%%%%%%%%%%%%%
%%%%%%%%%%%%%%%%% FIGURE 20 %%%%%%%%%%%%%%%%%%%
%%%%%%%%%%%%%%%%% FIGURE 22 %%%%%%%%%%%%%%%%%%%
%%%%%%%%%%%%%%%%% FIGURE 22 %%%%%%%%%%%%%%%%%%%
However, strong pressure dependence of the cyclotron effective
mass is not observed in de Haas-van Alphen experiments in
CeRhSi$_3$ \cite{terashima}. This is consistent with the result
of the less obvious pressure dependence of the resistivity
coefficient $A$. A small enhancement of the effective mass is
also suggested by heat-capacity measurements in CeIrSi$_3$
\cite{tateiwa2}.

In order to explain the positive curvature of $H_{c2}(T)$ and
the strong pressure dependence of $H_{c2}(0)$, Tada {\it et
al.} considered the temperature and pressure dependencies of
the correlation length of the spin fluctuations, $\xi_{sf}$,
\cite{tada2}. Since $\xi_{sf}$ is expressed as
$\xi_{sf}(T)=\frac{\tilde{\xi}_{sf}}{\sqrt{T+\theta}}$, in
which $\theta \rightarrow 0$ toward the QCP and the effective
pairing interaction is quadratically proportional to
$\xi_{sf}$, the superconducting coherence length $\xi(0)$ is
strongly reduced and $H_{orb}$ enhanced at low temperatures.
This model can explain the enhancement of the initial slope and
is compatible with the weaker enhancement of the effective
mass.

%--------------------------------------------------------------------------
\subsubsection{Superconducting Phase Diagram for Field in the Basal Plane}
In contrast to the high $H_{c2}(T)$ for the $c$ axis,
$H_{c2}(T)$ in the basal plane is not too high. However, it
still significantly exceeds $H_P^{\rm BCS}$ and is situated in
the upper part of Fig.~\ref{fig:pauli}. As discussed in the
section above, the high $H_{c2}(0)$ is attributed to the
reduced paramagnetic pair-breaking effect which originates
mainly from a non-vanishing spin susceptibility. In addition to
this, Agterberg {\it et al.} pointed out that another
characteristic mechanism, the helical vortex state, can also
evade the paramagnetic pair-breaking effect when the magnetic
field is applied in the basal plane
\cite{agterberg1,agterberg2}.

%%%%%%%%%%%%%%%%% FIGURE 24 %%%%%%%%%%%%%%%%%%%
\begin{figure}[]
\sidecaption 	
\includegraphics[width=0.5\textwidth]{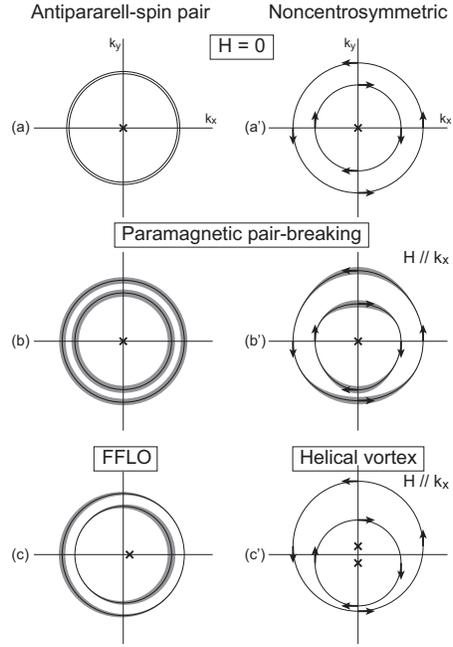}
\caption{Schematic illustrations of Fermi surface sliced
perpendicular to the $k_z$ axis. Pair breaking takes place at
the shaded region. The x marks indicate the positions of the
center-of-mass momenta of the Cooper pairs. } \label{fig:FFLO}
\end{figure}
%%%%%%%%%%%%%%%%% FIGURE 24 %%%%%%%%%%%%%%%%%%%

%Since details of the helical vortex state are given in the Chapter by D. F. Agterberg, here we focus on the difference between the helical vortex state, the paramagnetic pair-breaking effect and the Fulde-Ferrel-Larkin-Ovchinnikov (FFLO) state.

As discussed above, the paramagnetic pair-breaking effect
operates partially, with the center-of-mass momentum of the
Cooper pair remaining zero at any $k$ as shown in
Fig.~\ref{fig:FFLO}(b). On the other hand, in the helical
vortex state, the Fermi surfaces shift toward opposite
directions perpendicular to the field direction. An application
of a magnetic field does not break the Cooper pairs for all
$k$. In this case, the center-of-mass momenta of the Cooper
pairs belonging to each Fermi surface acquire a finite value
$\pm q$ (Fig.~\ref{fig:FFLO}(c)). The sign of $q$ depends on
the Fermi surface. In the Fulde-Ferrel-Larkin-Ovchinnikov
(FFLO) state, a Cooper pair with a finite center-of-mass
momentum is also realized. However, the pairing takes place in
a limited region on the Fermi surface. In other regions,
indicated by the shades in Fig.~\ref{fig:FFLO}(c), pairing is
not allowed. Since both states, helical vortex and FFLO phase,
evade the paramagnetic pair breaking, relatively high
$H_{c2}(0)$ can be realized.

\begin{figure}[]
\sidecaption 	
\includegraphics[width=0.6\textwidth]{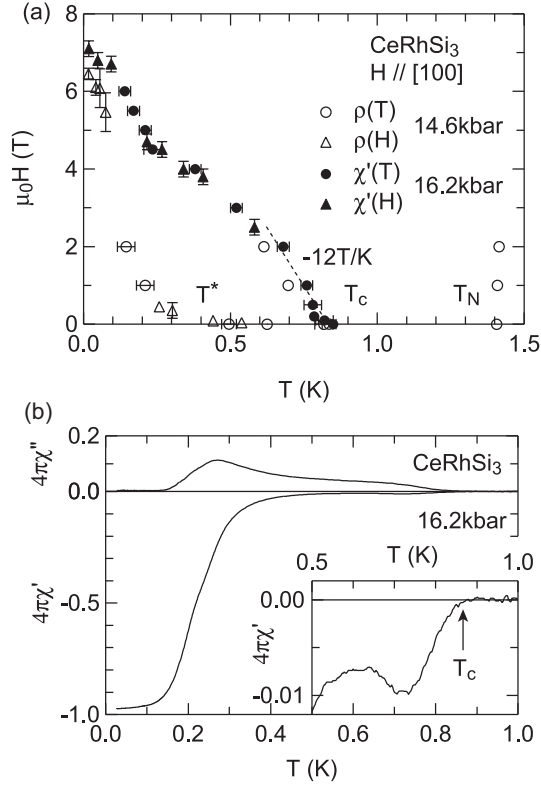}
\caption{(a) Magnetic field-temperature phase diagram of
CeRhSi$_3$ for $H\parallel a$ below $P^*_1$. $T^*$ denotes the
temperature at which an anomaly is observed in the resistivity
and magnetic ac susceptibility measurements. (b) ac
susceptibility as a function of temperature at zero field below
$P^*_1$. An enlarged view of the vicinity of $T_c$ is shown in
the inset \cite{kimura1}. } \label{fig:crs_pd_a}
\end{figure}

Although thus far no direct evidence for a helical vortex state
has been detected in either Ce$TX_3$ compounds or CePt$_3$Si,
some unusual superconducting properties for the field in the
basal plane are reported in CeRhSi$_3$ \cite{kimura1}. First,
there is a concave shape of the $H_{c2}(T)$ curve as shown in
Fig.~\ref{fig:crs_pd_a}(a).
%%%%%%%%%%%%%%%%% FIGURE 25 %%%%%%%%%%%%%%%%%%%
%%%%%%%%%%%%%%%%% FIGURE 25 %%%%%%%%%%%%%%%%%%%
The rapid change of $H_{c2}$ at low temperatures looks similar
to the one observed in a theoretically predicted helical-vortex
phase diagram \cite{agterberg2}. However, this feature observed
at $P<P^*_1$ becomes less obvious at  $P^*_3 \approx 2.6$ GPa
\cite{kimura4}. An influence of the antiferromagnetic order to
explain this unusual curve shape cannot be excluded. Second,
the ac susceptibility in the superconducting state shows an
unusual shape especially below $P^*_1$. The temperature at
which a large drop occurs in the real part of the
susceptibility $\chi^{\prime}$ is far below the onset
temperature of superconductivity. This might indicate that
superconductivity develops gradually in the antiferromagnetic
state. On the other hand, the imaginary part $\chi"$, namely
the energy dissipation associated with the dynamics of the
superconducting flux, is large even above the temperature at
which the large drop occurs in $\chi^{\prime}$. This
contradicts the view of a gradual development of
superconductivity. Although the influence of antiferromagnetism
is unclear at present, this rare behavior of the flux may be a
key feature to verify the helical vortex state.

\subsubsection{Energy Gap Structure and Pairing Symmetry}

\begin{figure}[]
\sidecaption 	
\includegraphics[width=0.6\textwidth]{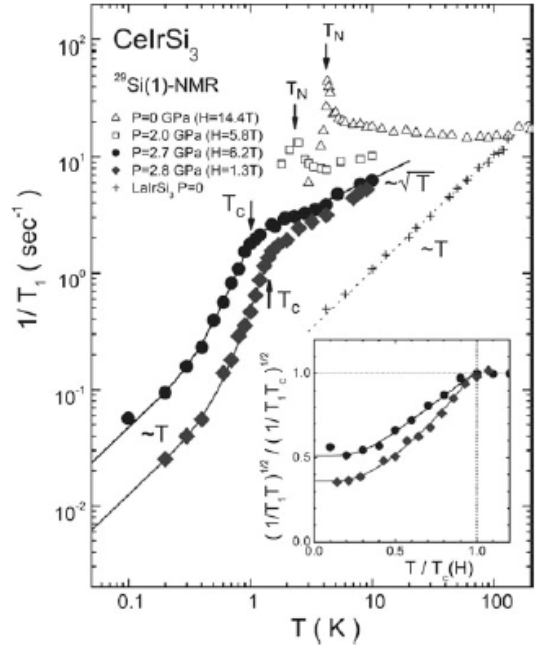}
\caption{Temperature dependence of $1/T_1$ measured by Si NMR
for CeIrSi$_3$ at several pressures. The solid curves below
$T_c$ for CeIrSi$_3$ indicate the calculated values obtained by
the line-node gap model with $2\Delta_0/(k_BT_c)\approx 6$ and
the residual density-of-states fraction $N_{\rm res}/N_0\approx
0.37$ (0.52) in $H=1.3$ T (6.2 T). The inset shows the plot of
$\sqrt{1/(T_1T)}$ normalized by that at $T_c$ which allows us
to evaluate $N_{\rm res}/N_0$ in the low-temperature limit
\cite{mukuda}.} \label{fig:nmr}
\end{figure}

In CeIrSi$_3$ the nuclear spin-lattice relaxation rate $1/T_1$
shows a $T^3$-like dependence below $T_c$ without a coherence
peak, as shown in Fig.~\ref{fig:nmr} \cite{mukuda}.  The data
are well fitted by the line-node gap model $\Delta=\Delta_0\cos
2\theta$. The fit yields $2\Delta_0/k_{\rm B}T_c\approx 6$,
which is much larger than the BCS weak-coupling value 3.53 and
thus suggests strong-coupling superconductivity. Using an
extended $s + p$ pairing state within a recent theoretical
model a behavior indicative of a line-node gap can be predicted
\cite{tada}.
%%%%%%%%%%%%%%%%% FIGURE 13 %%%%%%%%%%%%%%%%%%%
%%%%%%%%%%%%%%%%% FIGURE 13 %%%%%%%%%%%%%%%%%%%
This $1/T_1$-NMR measurement is the only one carried out to
determine the energy gap structure in Ce$TX_3$ compounds. To
identify the pairing symmetry in these materials, other
measurements, like the Knight shift, are highly desirable.

\subsection{Outlook}
The high anisotropy and strong enhancement of $H_{c2}$ seem to
be unique to Ce$TX_3$ superconductors. Interestingly, other HF
and non-HF superconductors without inversion symmetry do not
show these properties. The high orbital limiting field inherent
to the non-centrosymmetric HF Ce$TX_3$ superconductors
discloses the absence of the paramagnetic pair-breaking effect.
Conversely, absence of the effect unveils the unconventional
nature of the upper critical fields probably associated with
the quantum criticality of magnetism. Ce$TX_3$ compounds have
the potential to provide a vital clue to the underlying
mechanism of superconductivity mediated by magnetic
fluctuations.

To consider the relation between magnetism and
superconductivity in Ce$TX_3$, we need to keep in mind that the
magnetic ground states of CeCoGe$_3$ and CeIrGe$_3$ are
different from that of CeRhSi$_3$ and probably of CeIrSi$_3$.
CeCoGe$_3$ seems to display localized {\em f}-electron
magnetism, while CeRhSi$_3$ probably exhibits
itinerant-electron magnetism. It is very interesting that in
these compounds the $H_{c2}$ behaviors are similar in spite of
their different magnetism. The mass enhancement at $P^*_3$ is
suggested only in CeCoGe$_3$. The comparison of the
superconducting properties of CeCoGe$_3$ and CeIrGe$_3$ with
those of CeRhSi$_3$ and CeIrSi$_3$ will be important to
elucidate the nature of HF superconductors. Identification of
the gap structure in each compound is also a challenging issue,
which could provide possible evidence for the parity mixing of
the superconducting wavefunction. Moreover, some theoretically
predicted phenomena, like a helical vortex phase and a novel
magnetoelectric effect, remain to be verified in the future.

\section{UIr}

Whereas CeRhSi$_3$, CeIrSi$_3$, CeCoGe$_3$, CeIrGe$_3$ and
CePt$_3$Si are $4f$-electron antiferromagnets, UIr is a
$5f$-itinerant-electron ferromagnet with a Curie temperature
$T_{c1}=46$ K at ambient pressure \cite{dommann}. The
superconducting state in UIr appears to develop within a higher
pressure ferromagnetic phase at a critical temperature
$T_c=0.14$ K in a narrow pressure region around 2.6 GPa
\cite{akazawa1}. UIr is a moderate heavy-fermion compound with
a cyclotron mass $m*\sim 10-30~m_0$ \cite{yamamoto1}. The
coexistence of superconductivity and ferromagnetism imposes
several theoretical challenges, such as the mechanism and the
state of pairing. The pairing state in superconducting
ferromagnets needs to be spin triplet, otherwise the internal
exchange field would break the Cooper pairs. On the other hand,
a ferromagnetic state has a broken time reversal symmetry. The
superconducting BCS ground state is formed by Cooper pairs with
zero total angular momentum. The electronic states are
four-fold degenerate: $|\textbf{k}\uparrow\rangle$,
$|-\textbf{k}\uparrow\rangle$, $|\textbf{k}\downarrow\rangle$
and $|-\textbf{k}\downarrow\rangle$ have the same energy
$\epsilon(\textbf{k})$. The states with opposite momenta and
opposite spins are transformed to one another under time
reversal operation $\hat{K}|\textbf{k}\uparrow\rangle =
|-\textbf{k}\downarrow\rangle$, and the states with opposite
momenta are transformed to one another under inversion
operation $\hat{I}|\textbf{k}\downarrow\rangle =
|-\textbf{k}\downarrow\rangle$. The four degenerate states are
a consequence of spatial and time inversion symmetries. Parity
symmetry is irrelevant for spin-singlet pairing, but is
essential for spin-triplet pairing. Time reversal symmetry is
required for spin-singlet configuration, but is unimportant for
spin-triplet state \cite{anderson1,anderson2}. In UIr the lack
of spatial and time inversion symmetries lifts the degeneracies
and, therefore, superconductivity is not expected to occur.
Thus, UIr differs from the other two known ferromagnetic
superconductors UGe$_2$ \cite{saxena1} and URhGe \cite{daoki1},
in which the spatial inversion symmetry allows degeneracy in
the spin-triplet states. Theoretically and experimentally UIr
is a very special and challenging superconductor.

\subsection{Crystal Structure and Characteristic Parameters}
\label{subsec:21}

\begin{figure}
% \sidecaption
 \centering\scalebox{0.4}{\includegraphics{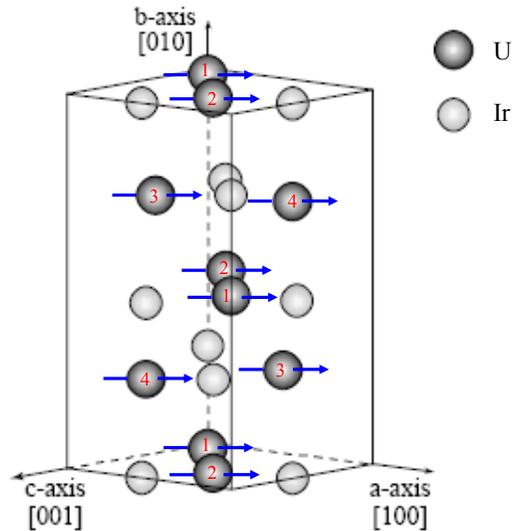}}
 \caption{Crystal and magnetic structures of UIr.}
 \label{f:1}
 \end{figure}

UIr crystallizes in a monoclinic PbBi-type structure (space
group $P2_1$) without inversion symmetry \cite{dommann}. The
lattice parameters are given in Table~\ref{tab:1}. The unit
cell has eight formula units with four inequivalent U and Ir
sites. The absence of inversion symmetry comes from the missing
mirror plane $(0,\frac{1}{2},0)$ perpendicular to the $b$ axis
(see Fig.~\ref{f:1}). Magnetism is of the Ising type with the
ordered magnetic moment oriented along the spin easy axis
[10$\bar{1}$] (Fig.~\ref{f:1}).

Single crystals of UIr have been grown by the Czochralski
method in a tetra-arc furnace
\cite{yonuki1,kobayashi1,yamamoto2,akazawa2,sakarya}. After
annealing, using the solid-state electrotransport technique
under high vacuum of the order of 10$^{-10}$~Torr, crystals
become of very high quality with residual resistivity $\rho_0
\sim 0.5~\mu\Omega$ and residual resistivity ratio (RRR)
$\rho_{300K}/\rho_0 \approx 200$ at ambient pressure.
Interestingly, single crystals of UIr seem to be of the highest
quality amongst those of non-centrosymmetric heavy-fermion
superconductors.

%
%--------------- Figure 1 -------------------
%------------- Figure 1 ----------------------
%
%

\begin{table}[htb]
\centering \caption{Normal and superconducting parameters of
$\rm UIr$} 	
\begin{tabular}{p{7.5cm}p{3cm}}
\hline\noalign{\smallskip}
		Crystal structure                              &   monoclinic  \\
		Space group                                    &   $ P2_1$        \\
		Lattice parameters                              &   $a=5.62$~{\AA} \\
		                            & $b=10.59$ {\AA} \\
                                    & $c= 5.60$ {\AA} \\
                                    & $\beta=98.9^\circ$ \\
\noalign{\smallskip}\hline\noalign{\smallskip}
		Sommerfeld value of specific heat              &   $\gamma_n = 40-49$~mJ/molK$^2$ \\
        Effective electron mass (Fermi sheet $\alpha$) &
        $m^* \sim 10-30~m_0$ \\
		Mean free path  (Fermi sheet $\alpha$)         &   $l = 1270$~{\AA}   \\
		Ferromagnetic transition temperature (ambient pressure)          &   $T_{c1}  =46$~K   \\
		Magnetic propagation vector                    &   $ \vec q = (1,0,-1)$  \\
        Magnetic moment $\mathbf{m_Q}$ along           &   [10$\bar{1}$] \\
		Saturated moment per U atom                    & $
\mu_s = 0.5~\mu_B$  \\
\noalign{\smallskip}\hline\noalign{\smallskip}
		Superconducting transition temperature         &   $T_c = 0.14$~K \\
		Upper critical field                           &   $H_{c2}(0) = 26$~mT \\
		Thermodynamic critical field                   &   $H_c(0) =  8$~mT \\
		Ginzburg-Landau coherence length               &   $\xi(0)= 1100$~{\AA}    \\
		Ginzburg-Landau parameter                      &   $\kappa \sim 2$  \\
\noalign{\smallskip}\hline\noalign{\smallskip}		
\end{tabular} 	
\label{tab:1}
\end{table}

\subsection{Normal State}
\label{subsec:22}

\subsubsection{Phase Diagram and Magnetic Properties}
\begin{figure}[b]
% \sidecaption
 \centering\scalebox{0.85}{\includegraphics{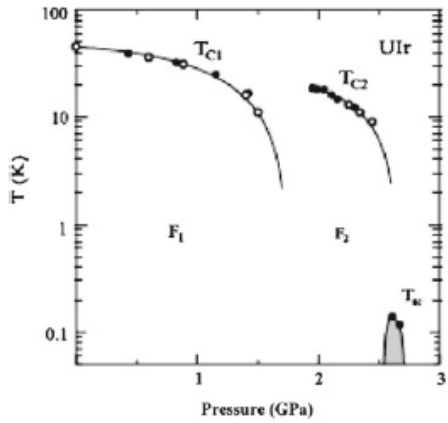}}
 \caption{Temperature-pressure phase diagram of UIr \cite{akazawa1}. F$_1$
 and F$_2$ are ferromagnetic phases.}
 \label{f:2}
 \end{figure}

Figure~\ref{f:2} shows the temperature-pressure phase diagram
as drawn by magnetization and resistivity measurements
\cite{akazawa1}.
%
%--------------- Figure 2 -------------------
%------------- Figure 2 ----------------------
%
%
The diagram consists of a low-pressure ferromagnetic phase FM1
(F$_1$ in the figure), a high-pressure ferromagnetic phase FM2
(F$_2$ in the figure) and a superconducting phase. A third
magnetic phase was reported \cite{kobayashi2}, but not further
evidence for it has been found. Application of pressure
decreases the Curie temperature $T_{c1}$ of the ferromagnetic
phase FM1 eventually to zero at the critical pressure $P_{c1}
\sim 1.7$~GPa. The FM2-paramagnetic curve appears just below 30
K and about 1.4 GPa, and goes away at a critical pressure
$P_{c3} \sim 2.7 -2.8$ GPa. Superconductivity is found in the
narrow pressure range 2.55-2.75 GPa below $T_c=0.14$ K.

\begin{figure}
 \centering
 \scalebox{0.9}{\includegraphics{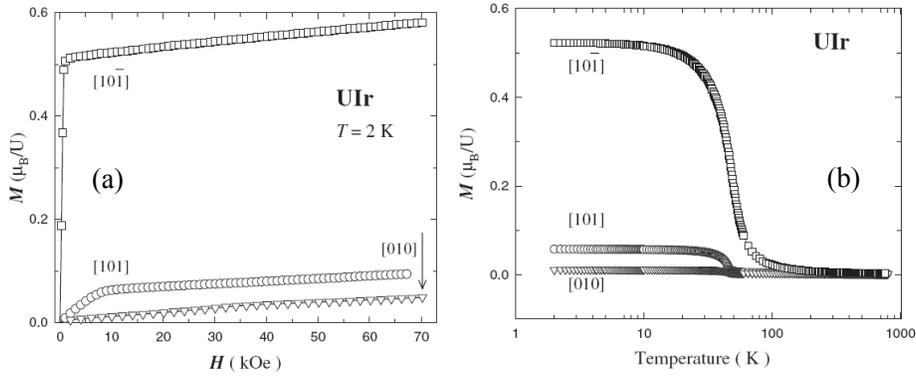}}
 \caption{Magnetization as a function of (a) field strength at 2 K and (b) temperature
 in 10 kOe in a single crystal of UIr \cite{galatanu1}.}
 \label{f:3}
 \end{figure}

The magnetic properties of this compound are governed by a
saturation moment along the easy axis [10$\bar{1}$], as
indicated by the magnetization curve of a single crystal at 2 K
shown in Fig.~\ref{f:3}(a) ~\cite{galatanu1}. The ordered
magnetic moment goes from 0.5$\mu_B$/U at ambient pressure in
the ferromagnetic phase FM1 to 0.07$\mu_B$/U at 2.4 GPa in the
ferromagnetic phase FM2 \cite{akazawa1}. The anisotropy of the
magnetization remains at high temperatures, as can be seen in
Fig.~\ref{f:3}(b). The susceptibility data follow a Curie-Weiss
law in the high-temperature paramagnetic region, with an
effective magnetic moment around 3.57~$\mu_B$/U that is pretty
close to the 5$f^2$ free-ion value 3.58~$\mu_B$/U. The small
value of the ordered moment 0.5$\mu_B$/U has been taken as
evidence for the itinerant character of the $5f$ electrons in
the ferromagnetic phase, though such a low value could also be
due to crystal-field effects \cite{sakarya,galatanu2}.

%
%
%--------------- Figure 2 -------------------
%------------- Figure 2 ----------------------
%
%

\subsubsection{Electronic States}
Quantum-oscillation and resistivity measurements provide
evidence that the low-temperature metallic state of UIr is a
Fermi liquid at ambient pressure. The dHvA measurements suggest
that the Fermi surface of UIr is two-dimensional and consists
mainly of nearly cylindrical sheets \cite{yamamoto1}. It has an
effective mass $m^* \sim 10-30~m_0$ and a mean free path $l
\sim 1270$~{\AA}. Such a value of the effective mass of the
$5f$ electrons leads to the classification of UIr as a moderate
heavy-fermion compound. Since the linear coefficient of the
heat capacity $\gamma_n \propto m^*$, summing for all the
branches yields the electronic specific-heat coefficient of
40-49~mJ/K$^2$mol \cite{yamamoto1,edbauer1}. There are no
band-structure calculations for this compound.

\begin{figure}
 \centering
 \scalebox{0.9}{\includegraphics{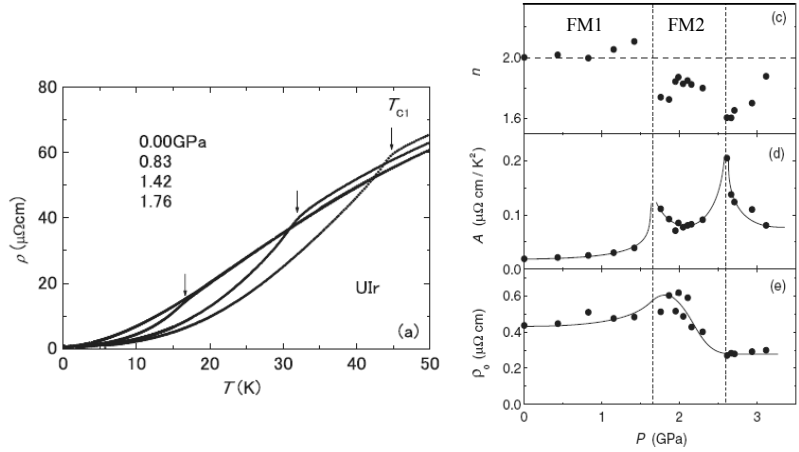}}
 \caption{(a) Resistivity as a function of temperature of UIr at different pressures,
 and (c-e) variations in the parameters $n$, $A$ and
$\rho_0$ of $\rho=\rho_0+AT^n$ as pressure increases
\cite{galatanu1}.}
 \label{f:4}
 \end{figure}

Figure~\ref{f:4}(a) shows the variation of the electrical
resistivity of UIr as a function of pressure
\cite{akazawa2,edbauer1}. At low pressure, in the ferromagnetic
phase FM1, the resistivity follows $\rho=\rho_0+AT^n$, with
$n\sim 2$ suggesting Fermi-liquid behavior. However, as
pressure increases the behavior becomes non-Fermi-liquid like
and eventually superconductivity appears in this regime.
Figures~\ref{f:4}(c-e) show the variation in $n$, $A$ and
$\rho_0$ as pressure increases. The non-Fermi-liquid behavior
above $\sim$ 1 GPa may be related to critical fluctuations near
the different magnetic transitions.

%
%--------------- Figure 2 -------------------
%------------- Figure 2 ----------------------
%
%

\subsection{Superconducting State}
\label{subsec:23}

\begin{figure}
 \centering
 \scalebox{0.8}{\includegraphics{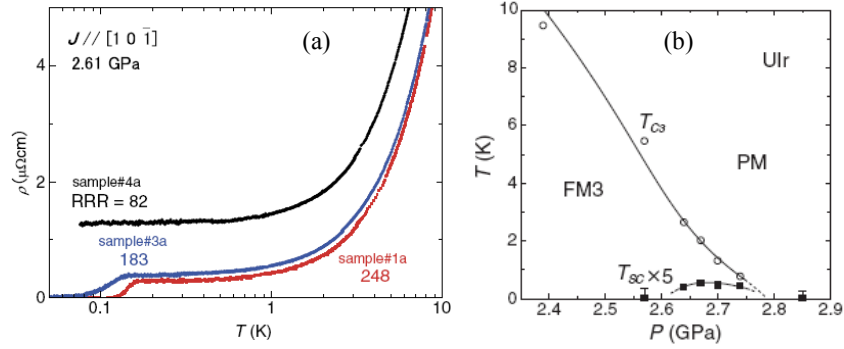}}
 \caption{(a) Resistivity below 10 K and at 2.61 GPa of UIr showing a
non-Fermi-liquid behavior. The three samples possess different
RRR-values: 82, 183 and 248. (b) Low-temperature region of the
phase diagram of UIr \cite{kobayashi2}.}
 \label{f:5}
 \end{figure}

Because of its extremely low critical temperature $T_c=0.14$~K
(in most figures in this section this critical temperature is
called $T_{sc}$), there is little information on the
superconducting state of UIr. Superconductivity seems to occur
inside and near the quantum critical point of the FM2 phase, in
the very narrow pressure range of 2.6-2.75 GPa \cite{akazawa1}.
In the ferromagnet UGe$_2$ with inversion symmetry the
superconducting phase exists inside the ferromagnetic phase as
well. Figure~\ref{f:5}(a) shows the temperature dependence of
the resistivity below 10 K and at 2.61 GPa, where it follows a
non-Fermi-liquid form $T^{1.6}$ \cite{akazawa2}.
Figure~\ref{f:5}(b) is a close-up of the low-temperature region
of the phase diagram where superconductivity appears (in this
figure, FM3 denotes the ferromagnetic phase FM2).

%
%--------------- Figure 2 -------------------
\begin{figure}
 \sidecaption
 \scalebox{0.6}{\includegraphics{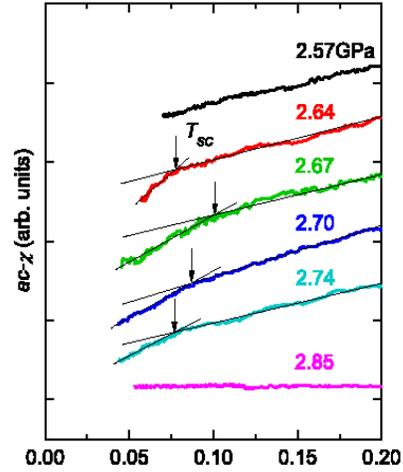}}
 \caption{Temperature dependence of the ac susceptibility of UIr below 0.2 K at
 different pressure. Arrows indicate the onset temperatures \cite{kobayashi2}.}
 \label{f:6}
 \end{figure}

%------------- Figure 2 ----------------------
%
%
Up to now experimental indications of the existence of a
superconducting phase in UIr come from measurements of
resistivity in samples of different qualities. No definite
diamagnetic signal has yet been observed in UIr
(Fig.~\ref{f:6}) \cite{kobayashi2}. The temperature dependence
of the resistivity for three different samples is shown in
Fig.~\ref{f:5}(a). The data indicate that superconductivity
becomes weaker as the residual resistivity ratio (RRR) of the
samples drops. Such a strong suppression of $T_{c}$ with
increasing impurities/defects is typical of unconventional
parity-conserving superconductors \cite{mineev}. Recent
theoretical works \cite{frigeri2,mineev2} considered impurity
effects on the critical temperature of superconductors without
inversion symmetry. It was found that impurity scattering leads
to a functional form of $T_c$ that, up to a prefactor, is the
same as the one for unconventional superconductor with
inversion symmetry: $\ln(T_c/T_{c0}) =
\alpha\left[\Psi(\frac{1}{2})-\Psi\left(\frac{1}{2}-\frac{\Gamma}{2\pi
T_c}\right)\right]$. The suppression of $T_c$ by impurities in
non-centrosymmetric UIr agrees with this prediction
\cite{frigeri2,mineev2}.

%
%--------------- Figure 2 -------------------
%------------- Figure 2 ----------------------
%
%

\begin{figure}
 \sidecaption
 \scalebox{0.5}{\includegraphics{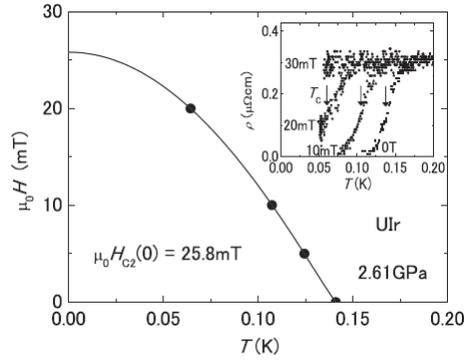}}
 \caption{Upper critical field $H_{c2}(T)$ along the
easy axis [10$\bar{1}$] in UIr. The critical temperatures were
defined by the midpoints of the resistivity drops in the inset
\cite{akazawa2}.}
 \label{f:7}
 \end{figure}

The upper critical field $H_{c2}(T)$ in the direction of the
easy axis [10$\bar{1}$] in the high-temperature region was
determined by resistivity measurements in a sample with a very
high RRR (Fig.~\ref{f:7}) \cite{akazawa2}. $T_{c}$ was defined
as the midpoint of the resistivity drop. By assuming the
standard empirical expression
$H_{c2}(T)=H_{c2}(0)\left[1-(T/T_{c})^2\right]$, the
zero-temperature upper critical field $H_{c2}(0)$ was estimated
as 26~mT corresponding to a coherence length of $\xi(0)=
1100$~{\AA}.  Since this value of $H_{c2}(0)$ is smaller than
the paramagnetic limiting field $H_p=280$~mT orbital depairing
is the likely mechanism for the upper critical field in UIr.

%
%--------------- Figure 2 -------------------
%------------- Figure 2 ----------------------
%
%

It is believed that near a ferromagnetic quantum critical point
spin fluctuations lead to Cooper pairing in a spin-triplet
channel. There are some possible scenarios for the unexpected
realization of superconductivity in this compound. It needs to
be confirmed that FM2 is indeed a ferromagnetic phase, and not a
canted antiferromagnetic phase that could yield pairing in the
spin-single channel. A canted antiferromagnetic phase may be
caused by the spin-orbit coupling and the low symmetry of the
crystalline structure without inversion symmetry, as discussed
by Dzyaloshinky and Moriya. In this sense, it is important to
note that the saturated moment of 0.07~$\mu_B$/U in the FM2
phase is quite small. Another possibility is the
FFLO state, in which at zero
magnetic field electrons with momenta \textbf{k} and
\textbf{-k+q} can pair with nonzero angular momenta. A
mean-field model has been recently proposed for
superconductivity in non-centrosymmetric ferromagnets
\cite{linder1}, in which the antisymmetric spin-orbit coupling
(ASOC) turns out to enhance both superconductivity and
ferromagnetism in all spin channels. Future experiments will be
absolutely important for the understanding of this unique
superconductor.

\section{Comparison of the Superconducting
States of CePt$_3$Si, CeRhSi$_3$, CeIrSi$_3$, CeCoGe$_3$, CeIrGe$_3$ and
UIr}

The superconducting properties of the non-centrosymmetric HF
compounds have not been easy to determine. On the one hand, the
Ce$TX_3$ and UIr compounds only superconduct under pressure,
making technically difficult to study them. On the other hand,
CePt$_3$Si does become superconducting at ambient pressure, but
has drawbacks in the crystal quality available. In spite of
this, it has been possible to establish some of the
characteristics of these materials.

The Ce$TX_3$ materials, with tetragonal space group $I4mm$, and
CePt$_3$Si, with tetragonal space group $P4mm$, have the same
generating point group $C_{4\nu}$ which lacks a mirror plane
and a two-fold axis normal to the $c$ axis. Thus, a Rashba-type
interaction appears in all these compounds due to the missing
of inversion symmetry. In contrast, UIr has a monoclinic
lattice.

The Sommerfeld coefficient is much larger in CePt$_3$Si than in
Ce$TX_3$ and UIr. Electron correlations should hence be
stronger in CePt$_3$Si. On the other hand, all compounds turn
from Fermi-liquid to non-Fermi-liquid states as pressure is
increased and become superconducting in the Fermi-liquid state,
with the exception of UIr.

The behavior of the upper critical field $H_{c2}$ in the
Ce$TX_3$ systems is very different from that in CePt$_3$Si. In
Ce$TX_3$, $H_{c2\parallel c}> 22$~T and $(-dH_{c2\parallel c
}(T)/dT)_{T_c}>17$~T/K are larger than $H_{c2\parallel c}
\approx 3$~T and $(-dH_{c2\parallel c}(T)/dT)_{T_c}\sim6.3$~T/K
in CePt$_3$Si. Moreover, in Ce$TX_3$ $H_{c2}(T)$ has a positive
curvature, unlike in CePt$_3$Si. In CeIrSi$_3$, for example,
superconductivity is anisotropic ($H_{c2\parallel
c}/H_{c2\parallel ab})> 3$, whereas in CePt$_3$Si it is almost
isotropic. There are clear signatures for unconventional
superconductivity in CeIrSi$_3$, CeRhSi$_3$ and CePt$_3$Si,
including evidence for line nodes in some cases. The fact that
these compounds also support strong magnetic features and order
suggests strongly that unconventional pairing mechanisms could
be at work here. In this context it is particularly intriguing
to analyze the role the antisymmetric spin-orbit coupling.

\begin{acknowledgement}
We are grateful to M. Sigrist, S. Fujimoto, Y. Tada, D. F.
Agterberg, K. Samokhin, F. Honda, Y. Matsuda, T. Sugawara, T.
Terashima, H. Aoki, F. L\'evy and I. Sheikin for helpful
discussions. The work of N.K. was supported by KAKENHI (grant
numbers 18684018 and 20102002) and partially by MEXT of Japan
through Tohoku University GCOE program ``Weaving Science Web
beyond Particle-matter Hierarchy". I.B. acknowledges initial
support by the Venezuelan FONACIT (grant number S1-2001000693).
\end{acknowledgement}

%\bibliographystyle{spphys}
%\bibliography{NCSSC_book}

\end{document}